\documentclass[12pt]{article}

\usepackage{epsfig}

\newcount\xrefpos \xrefpos=0
\newcount\yrefpos \yrefpos=0
\newcount\xput \xput=0
\newcount\yput \yput=0

\def\refpos#1 #2 #3{\global\xrefpos=#1 \global\yrefpos=#2
                         \rlap{$\smash{#3}$}}
\def\put #1 #2 #3{\xput=#1 \yput=#2
                  \advance\xput by -\xrefpos
                  \advance\yput by -\yrefpos
                  \rlap{\kern\the\xput truebp
                        \vbox to 0pt{\vss\hbox{$\displaystyle #3$}
                        \kern\the\yput truebp}}}
\def\beginlabels\refpos#1\endlabels{\hbox{$\refpos#1$}}

\usepackage{amsfonts,amssymb}

\newcommand{\sect}[1]{\setcounter{equation}{0}\section{#1}}

\newcommand{\al}{\ensuremath{\alpha}}

\newcommand{\vep}{\ensuremath{\varepsilon}}
\newcommand{\ka}{\ensuremath{\kappa}}
\newcommand{\la}{\ensuremath{\lambda}}

\newcommand{\Om}{\ensuremath{\Omega}}
\newcommand{\p}{\ensuremath{\phi}}

\newcommand{\Q}{\ensuremath{{\cal Q}}}

\renewcommand{\d}{\ensuremath{{\rm d}}}

\newcommand{\ra}{\ensuremath{\rightarrow}}
\newcommand{\del}{\ensuremath{\partial}}

\newcommand{\td} {\ensuremath{\tilde}}

\newcommand{\be}{\begin{equation}}
\newcommand{\ee}{\end{equation}}
\newcommand{\ba}{\begin{eqnarray}}
\newcommand{\ea}{\end{eqnarray}}

\begin{document}

\begin{titlepage}

\bigskip
\hskip 4.8in\vbox{\baselineskip12pt \hbox{hep-th/0410214}
\hbox{NSF-KITP-04-118}}

\bigskip
\bigskip
\bigskip

\begin{center}

{\Large \bf On Charged Black Holes in Anti--de
Sitter Space}
\end{center}

\bigskip
\bigskip
\bigskip

\centerline{{\bf Dominic Brecher}$\,^a$, {\bf Jianyang He}$\,^a$ {\bf and
Moshe Rozali}$^{\,a,b}$}

\bigskip
\bigskip
\bigskip

\centerline{$^a$ \it Department of Physics and Astronomy}
\centerline{\it University of British Columbia}
\centerline{\it Vancouver, British Columbia V6T 1Z1, Canada}

\vskip0.5cm

\centerline{$^b$ \it Kavli Institute for Theoretical Physics}
\centerline{\it University of California}
\centerline{\it Santa Barbara, CA 93106-4030, USA}

\vskip0.5cm

\centerline{\small \tt
brecher, jyhe, rozali@physics.ubc.ca}

\bigskip
\bigskip

\begin{abstract}
\vskip 2pt We study the region inside the event horizon of charged
black holes in five dimensional asymptotically anti-de Sitter space,
using as a probe two-sided correlators which are dominated by
spacelike geodesics penetrating the horizon. The spacetimes we
investigate include the Reissner-Nordstr\"{o}m black hole and
perturbations thereof. The perturbed spacetimes can be found
exactly, enabling us to perform a local scan of  the region between
the inner and outer horizons. Surprisingly, the two-sided
correlators we calculate seem to be geometrically protected from the
instability of the inner horizon.

\end{abstract}

\end{titlepage}


\baselineskip=18pt \setcounter{footnote}{0}

\sect{Introduction and Summary}

Generic black holes, those with non-zero charge and/or
angular momentum, have a causal structure
very different to that of the more familiar Schwarzschild black hole.
This is a source of new and interesting phenomena.  In addition to the
outer event horizon, the geometry contains an inner horizon
surrounding a singularity, which is timelike. This
inner horizon is also a Cauchy horizon -- it bounds a
region outside of the Cauchy development of an
initial data set. Finally, the inner
horizon is a surface of infinite blue shift -- analysis at the linearised
level shows that
an arbitrary perturbation gets magnified by the
geometry near that surface~\cite{penrose:68,matzner,chan:82}.

It is then generally expected that the inner
horizon is unstable to small perturbations, which back-react strongly on the
geometry, likely resulting in a singularity. The
spacetime effectively ends at or before one reaches
the Cauchy horizon, and so the cosmic censorship
conjecture~\cite{penrose} is upheld. However, the
precise mechanism
for this expected instability, and the resulting
spacetime structure, have not yet been settled. For recent
efforts towards clarifying this issue within
classical general relativity see~\cite{dafermos}, and for a recent
investigation of three dimensional rotating black holes in this
context see~\cite{vijay}.

This set of issues could be studied within the AdS/CFT
correspondence~\cite{adscft,adscft2} although, as for other interesting
phenomena\footnote{See, however,
  recent efforts to construct naked singularities in asymptotically
  AdS spaces~\cite{naked}.} associated with black holes, to study the
physics one would have to probe the geometry behind an event horizon.
Early efforts to do that
are~\cite{simon:98,simon:99,simon:00,hubeny:02}.  Here, we
will use the set of ideas introduced in~\cite{juan} and further
developed in \cite{hemming,kraus:03,stanford,simon,vijay,kaplan,fidkowski}.
Namely, we will study a set of
observables which depend in a precise sense on the physics behind the
horizon\footnote{Though, in the spirit of black
hole complementarity~\cite{complement}, should also be recoverable from
observations outside the horizon only.}.

The observables are correlation functions of operators
which are inserted on the two disconnected
boundaries of the eternal black hole spacetime~\cite{juan}.  These are
the objects we will use to probe the above questions associated with generic
black holes.  In the limit of large operator dimension, the
correlators are approximated by
the contribution from a single spacelike geodesic,
which penetrates the event horizon.  For a precise
explanation of this approximation we refer the
reader to~\cite{juan,stanford}.

This is an explicit realisation of work by Israel~\cite{oldisrael},
in which parallels were drawn between the concepts of the
thermofield formalism of field theory (see, e.g.,~\cite{thermofield}
for a review) and black hole physics.  The fictitious Hilbert space
in the thermofield formalism is just the space of states behind the
event horizon of the eternal black hole. By tracing over the
fictitious states, we recover the thermal behaviour of the physical
states.  In the AdS version, correlators of operators inserted on a
single boundary will be ordinary thermal correlation functions, but
one may also consider correlation functions of operators inserted on
 two opposite boundaries. As explained
in~\cite{juan,stanford}, one can compute such correlators from those
of operators inserted on a single boundary, by  analytic
continuation, simply shifting the argument of one operator by half a
period of imaginary time.

The outline of the paper, and a summary of results is as follows:

We begin, in section 2, by introducing the five dimensional
AdS-Reissner-Nordstr\"{o}m (AdS-RN) geometry, following
\cite{RN}.  We discuss
coordinate systems and complexification of the
metric, and draw the Penrose diagram which encodes
the causal structure.  We should note that we will only be concerned
with the non-extremal black hole, since the extremal solution has only
one asymptotic region, so there are simply no such objects as
two-sided correlators.  This makes sense, since the extremal black
hole is at zero temperature~\cite{RN}.

In sections 3 and 4, we investigate correlation functions of neutral
and charged operators respectively.  The discussion does not involve
the subtleties of~\cite{stanford}, as there is no branch cut or
singularity, due to the drastically different causal structure.  We
find, however, an interesting non-analytic feature of the
correlation function, which exists only for black holes sufficiently
far from extremality.  It would be interesting to discover the
meaning of such a phase transition in the field theory language.

As emphasised in~\cite{simon}, it seems that the field theory only encodes the
region of spacetime within the Cauchy horizon.  In that sense, the AdS/CFT
correspondence would uphold the cosmic censorship conjecture, at least for
the observables we consider.  It would be interesting to see if this is a
more general feature of the field theory.

In section 5 we turn to an analysis of a perturbed AdS-RN spacetime.
For perturbations which move at the speed of light (ingoing or
outgoing only), one can use the methods of Poisson and
Israel~\cite{poisson} to find the perturbed metric exactly.  By
varying the time at which the metric is perturbed, one can perform a
scan of (at least some part of) the region between the inner and
outer horizons\footnote{This is somewhat similar to the discussion
in \cite{kaplan}, though here the perturbed metric is on-shell.},
again using the two-sided correlators as a probe.  We demonstrate
this through the simplest example of an infinitely thin
perturbation, an outgoing thin shell of null matter.

We confine ourselves here to calculating the dependence of the
correlators on the time at which the perturbation leaves the
boundary. This demonstrates that the gauge
theory is sensitive to a perturbation of  spacetime which is localised
purely behind the outer horizon. It would be interesting to read off
from these correlators the dramatic behaviour expected when mass
inflation \cite{poisson} sets in. We leave this to future work
\cite{progress}, but we comment that so far it seems that our
correlators are screened from this catastrophic instability by some
subtle geometrical effects, which we explain below.


\sect{Classical Geometry}

In this section we discuss the classical geometry of the electrically
charged AdS-RN black
hole, its complexification and causal structure.  We restrict
attention to aspects of the geometry which are relevant for us, more
details can be found in~\cite{RN}.


\subsubsection*{The Spacetime}

The five dimensional solution\footnote{A solution of five dimensional
  Einstein-Maxwell theory with a negative cosmological constant.  The
  ten dimensional origins are discussed in~\cite{RN}.} takes the
  form~\cite{RN} \be \d s^2 =
-f(r) \d t^2 + \frac{\d r^2}{f(r)} + r^2 \d \Om_3^2, \qquad A =
\left( -\frac{1}{c} \frac{Q}{r^2} + \Phi \right) \d t
\label{eqn:soln} \ee where \be f(r) = 1 + \frac{r^2}{l^2} -
\frac{M}{r^2} + \frac{Q^2}{r^4} \equiv \frac{\Delta(r)}{l^2r^4},
\qquad \Delta(r) = r^6 + l^2 r^4 - l^2 M r^2 + l^2 Q^2, \ee $c =
2/\sqrt{3}$, $\Phi$ is a constant and $M$ and $Q$ are respectively
proportional to the mass and charge of the black hole. The spacetime
is asymptotically AdS, with curvature radius $l$, the boundary
theory living on $\mathbb{R} \times S^3$.  From now on, we will take
$l=1$.  The boundary is at $r=\infty$ and there is a timelike
singularity at $r=0$.

Horizons of the metric are given by the real positive roots of
$\Delta(r)$.  We will often work with the variable $x=r^2$,
$\Delta(x)$ being a cubic in $x$.  Since, with $M > 0$, \be
\Delta(0) = Q^2 > 0 \qquad {\rm and} \qquad \frac{\d\Delta}{\d x}(0)
= -M < 0, \ee one root of $\Delta$ must occur for $x<0$.  We denote
this negative root by $-x_0 = -r_0^2$.  There are then at most two
positive real roots, denoted by $x_\pm = r_\pm^2$, with $x_+\ge
x_-$.  We will thus write \be \Delta(x) = (x+x_0)(x-x_-)(x-x_+). \ee
It is easy to show that $x_0=1+x_-+x_+$, so we will only need to
specify $x_\pm$.  The non-extreme black hole, with $x_+ > x_-$, has
an outer event horizon at $r_+$ and an inner Cauchy horizon at
$r_-$.  The extreme solution, with $x_+=x_-$, has a single event
horizon at $r_+$.

The surface gravities at the outer and inner horizons are,
respectively, \be \ka_\pm =
\frac{(x_0+x_\pm)(x_+-x_-)}{x_\pm^{3/2}}, \ee the temperature of the
black hole being $T = \ka_+/(2\pi) = 1/\beta$.  The dual field
theory is thus in a thermal state at temperature $T$.  One can then
work~\cite{RN} with either fixed charge (the canonical ensemble) or
fixed potential (the grand canonical ensemble).  In the latter case,
the electrostatic potential in the field theory is given by the
constant $\Phi$, which can be fixed as $\Phi = (1/c)(Q/x_+)$ such
that $A_t (r_+) = 0$. In terms of the bulk physics, $\Phi$ is the
electrostatic potential difference between the horizon and infinity.


\subsubsection*{Embedding into the Complex Plane}

There are six distinct regions of the spacetime, the Penrose
diagram consisting of an infinite sequence of these basic blocks.
As was emphasized in \cite{simon}, and as will be discussed below, the
region behind the inner horizon does not seem to play any role in
the field theory, therefore we can restrict attention to one copy
of these basic blocks.

To describe the global structure, one can pass to Kruskal-like
coordinates to describe the maximal analytic extension of the
space. Alternatively, as
in~\cite{kraus:03,stanford,simon,vijay}, one can use different
Schwarzschild coordinate patches to describe the global extension.
They are embedded into complex Schwarzschild time, with $t = t_L + i
t_E$ having a constant imaginary part in each patch.  This is
shown in figure 1, in which we define $t_E = 0$ in the right
asymptotic region. Then, as in~\cite{stanford}, crossing the event
horizon at $r_+$ shifts $t_E$ by $\pi/(2\ka_+) = i\beta/4$ so, to
move from a point on the right-hand boundary to the symmetric
point on the left-hand boundary, we shift $t \ra -(t+i\beta/2)$.
Crossing the Cauchy horizon at $r_-$ instead would shift $t_E$ by
$\pi/(2\ka_-)$ although this will not be relevant to the boundary
field theory\footnote{The embedding into
complex time can be repeated infinitely in both directions, by shifting
$t_E$ appropriately, as one crosses each horizon.}.

\begin{figure}[ht]
         \beginlabels\refpos 0 0 {}
                     \put 155 -60 {r<r_-}
                     \put 150 -80 {t_E = -\beta/4}
                     \put 160 -95 {+ \pi/(2\ka_-)}
                     \put 275 -60 {r<r_-}
                     \put 260 -80 {t_E = -\beta/4}
                     \put 270 -95 {-\pi/(2\ka_-)}
                     \put 205 -150 {r_- < r < r_+}
                     \put 205 -175 {t_E = -\beta/4}
                     \put 155 -240 {r>r_+}
                     \put 150 -260 {t_E = -\beta/2}
                     \put 265 -240 {r>r_+}
                     \put 265 -260 {t_E = 0}
                     \put 205 -300 {r_- < r < r_+}
                     \put 208 -320 {t_E = \beta/4}
         \endlabels
         \centerline{
         \psfig{figure=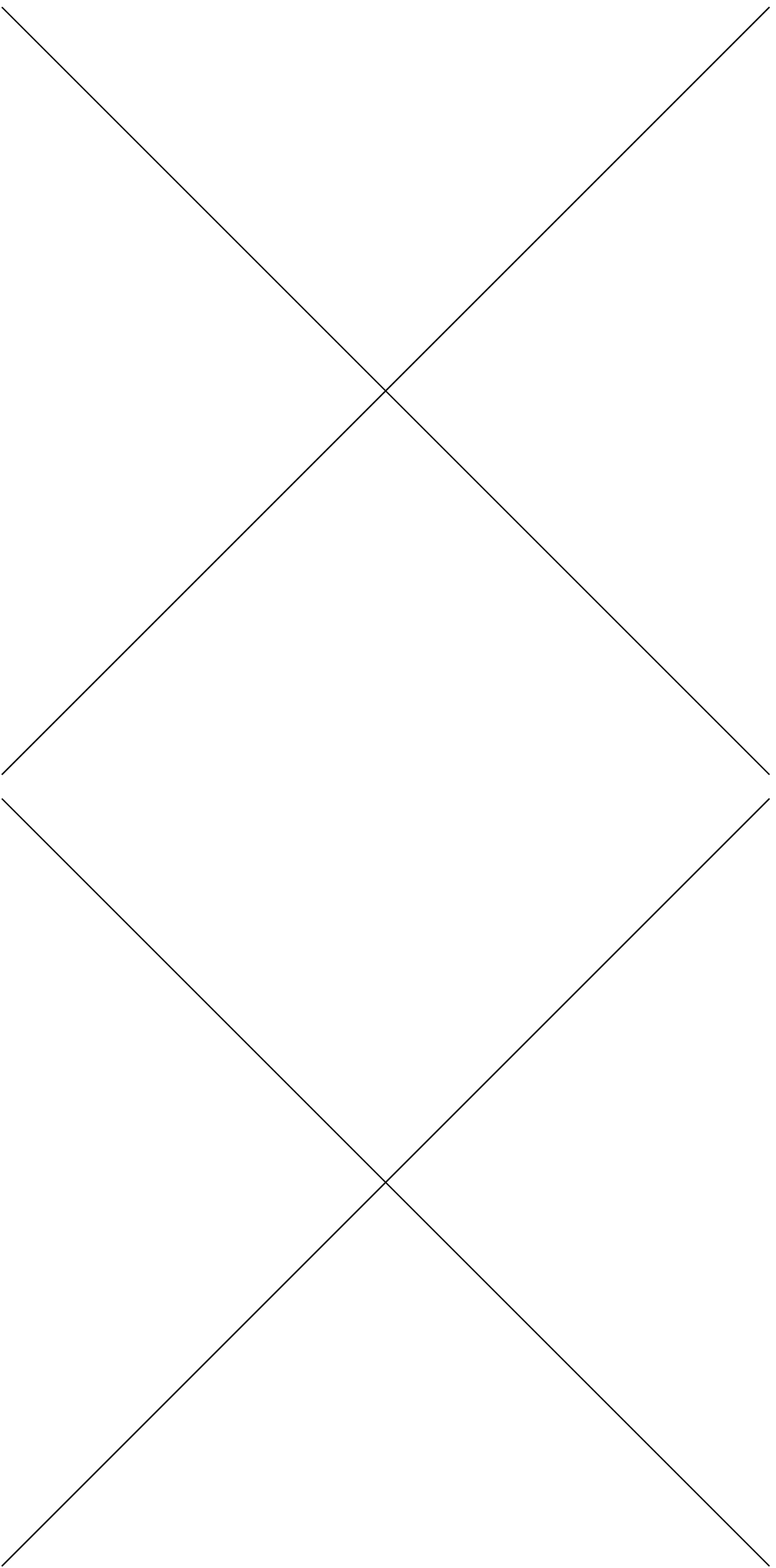,height=4.5in}}
         \vskip0.5cm
         {\footnotesize Figure 1.  The embedding into complex
                     Schwarzschild time, showing the constant imaginary parts
                     of $t=t_L + i t_E$.}
\end{figure}


\subsubsection*{The Penrose Diagram}

The basic features of the Penrose diagram are shown by the behaviour
of radial null geodesics. As in the uncharged case~\cite{stanford},
the boundaries and the singularities cannot both be drawn as
straight lines.  For radial null geodesics, we work with the
Lagrangian \be {\cal L} = - f(r) \dot{t}^2 + \frac{\dot{r}^2}{f(r)} = 0
\ee where dots denote differentiation with respect to the affine
parameter $\lambda$.  There is a conserved energy, $E$, associated
with the Killing vector $\del/\del t$, in terms of which the
geodesic equations become \be \dot{t} = \frac{E}{f(r)}, \qquad
\dot{r}^2 = E^2. \ee

For ingoing geodesics which start at the boundary $r=\infty$ at $t =
0$, the time $t$ as a function of $r$ is given by \be t(r) =
\int_r^{\infty} \frac{\d r'}{f(r')} = \int_{r^2}^{\infty} \frac{\d
x}{2} \frac{x^{3/2}}{(x+x_0)(x-x_-)(x-x_+)}. \ee Explicitly, for $r
< r_-$, \ba t(r) &=& - \frac{x_0^{3/2}}{(x_0+x_-)(x_0+x_+)}
\left( \tan^{-1} \left(\frac{r}{r_0}\right) - \frac{\pi}{2} \right)
- \frac{1}{\ka_-}
\tanh^{-1} \left(\frac{r}{r_-}\right) \nonumber \\
&& + \frac{1}{\ka_+} \tanh^{-1}
\left(\frac{r}{r_+}\right) -\frac{i\pi}{2}\left(
  \frac{1}{\ka_+} - \frac{1}{\ka_-} \right)
\label{eqn:t} \ea where the pure imaginary terms
arise from integrating over the two poles at
$x=x_\pm$.  These are just the shifts in $t_E$
discussed above.  Such a geodesic will reach the
singularity $r=0$ at a finite time
\be
t_{\rm sing} = \frac{\pi}{2}
\frac{x_0^{3/2}}{(x_0+x_-)(x_0+x_+)}
-\frac{i\pi}{2}\left(
  \frac{1}{\ka_+} - \frac{1}{\ka_-} \right).
\ee whose real part is positive. Conversely, a geodesic which hits
the singularity at $t_L = 0$, must leave the boundary at time \be
t_c = - \frac{\pi}{2} \frac{x_0^{3/2}}{(x_0+x_-)(x_0+x_+)} < 0, \ee
from which we conclude that one cannot draw both the boundaries and
the singularities as straight lines in the Penrose diagram. Choosing
the boundaries to be straight lines, then the singularities must be
\textit{bowed out}, the situation demonstrated in figure 2. However,
it will become clear that this behaviour has no influence on the
boundary theory. In particular the critical value $t_c$ in the
Schwarzschild case of~\cite{stanford} plays no role in the
discussion of correlation functions here.

\begin{figure}[ht]
         \beginlabels\refpos 0 0 {}
                     \put 145 -60 {t=\infty}
                     \put 140 -170 {t=-\infty}
                     \put 145 -270 {t=\infty}
                     \put 128 -90 {t=t_c}
                     \put 125 -112 {t=0}
                     \put 293 -60 {t=-\infty}
                     \put 293 -170 {t=\infty}
                     \put 293 -270 {t=-\infty}
                     \put 293 -220 {t=0}
                     \put 293 -248 {t=-t_c}\
         \endlabels
         \centerline{
         \psfig{figure=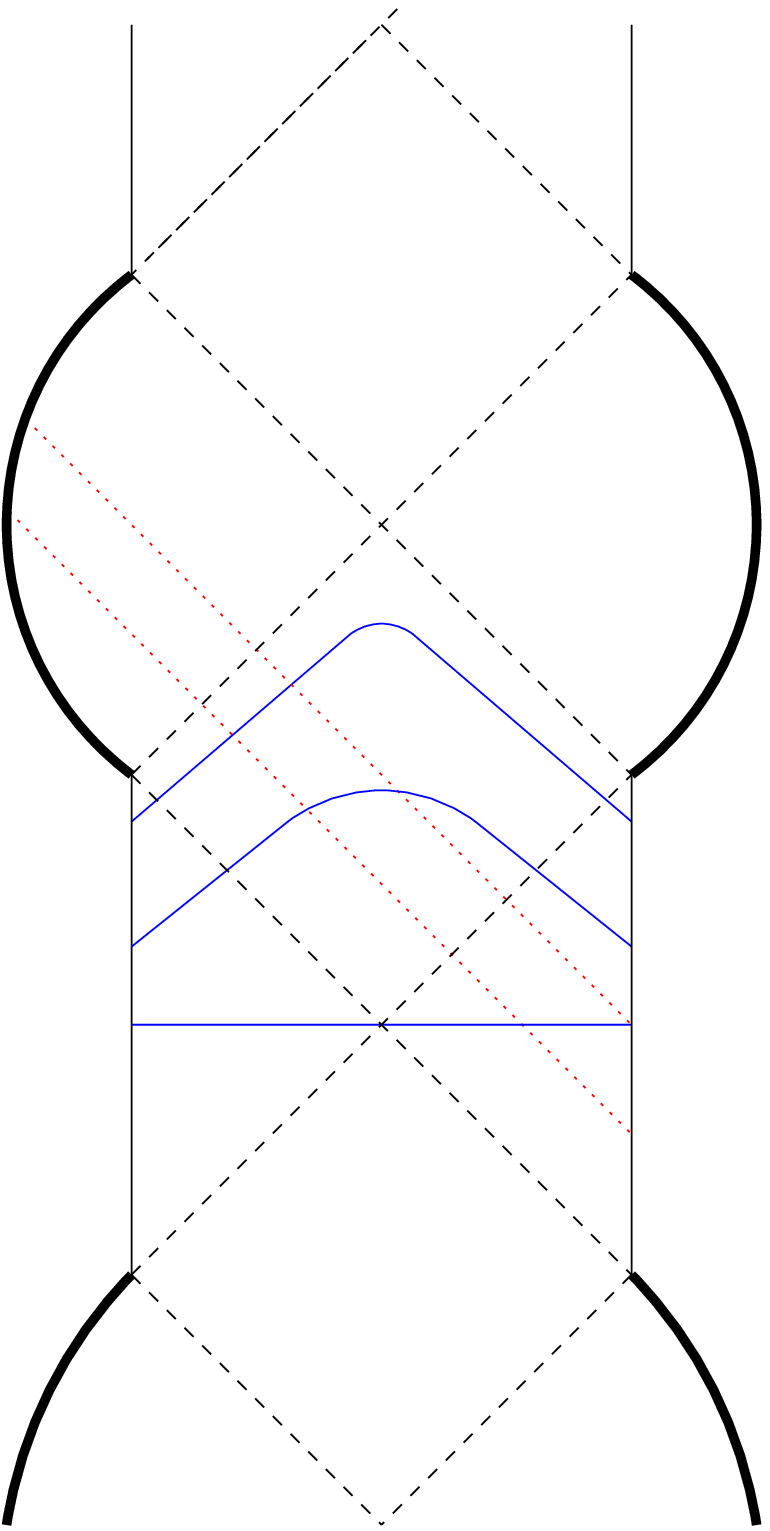,height=4.5in}}
         \vskip0.5cm
         {\footnotesize Figure 2.  The Penrose diagram repeats itself
         in both directions.  Radial null geodesics are dotted red,
                     radial spacelike geodesics are solid blue, the
                     inner and outer horizons are dashed and the
                     singularities are bold.  We have chosen to draw
                     the boundaries as
                     vertical lines, in which case the singularities
                     are bowed out.  The values of $t$ all denote
                     those of $t_L$, the real part of complex time.}
\end{figure}


\subsubsection*{Kruskal Coordinates}

As in the asymptotically flat case~\cite{chan:85},
we need two Kruskal-like coordinates patches to
cover the region $0<r<\infty$, one being valid
through the outer horizon, the other being valid
through the inner horizon.  In the usual way, we
first define the tortoise coordinate \be r_* =
\int_0^r \frac{\d r'}{f(r')} + C = x_0^\prime
\tan^{-1} \left(
  \frac{r}{r_0} \right) + \frac{1}{2\ka_+} \ln
  \left( \frac{r-r_+}{r+r_+} \right) - \frac{1}{2\ka_-} \ln
\left( \frac{r-r_-}{r+r_-} \right)
\label{eqn:rstar} \ee where $x_0^\prime =
x_0^{3/2}/((x_0+x_+)(x_0+x_-))$ and we have chosen
the constant $C$ so as to make $r_*$ real for
$r>r_+$.  As we cross the horizons from this
region, we pick up the same constant imaginary
terms as in the expression (\ref{eqn:t}) for $t$,
giving an embedding of the spacetime into the
complex coordinate plane.

To cover the entire range of $r$, define the light cone coordinates
$u=t-r_*,v=t+r_*$.  For $r>r_-$, transform according to
\be
U^+ =
-e^{-\ka_+ u} = T^+ - X^+, \qquad V^+ = e^{+\ka_+ v} = T^+ + X^+
\label{eqn:kruskal}
\ee
whereas for $r<r_+$, take
\be
U^- = -e^{+\ka_- u} = T^+ - X^+, \qquad V^- = e^{-\ka_- v} = T^- +
X^-.
\ee
The outer horizon is given by $T^{+2}-X^{+2} = 0$ and the inner has
$T^{-2}-X^{-2} = 0$.  The metric becomes \be \d s^2 =
\frac{1}{\ka_\pm^2} f(r) ~e^{\mp 2\ka_\pm r_*} (-(\d T^\pm)^2 + (\d
X^\pm)^2 ), \ee where for $r>r_-$, $r(T^+,X^+)$ is determined
implicitly by
\be
T^{+2}-X^{+2} = - e^{2\ka_+ r_*} = -e^{2\ka_+ x_0^\prime
  \tan^{-1}(r/r_0)} \left( \frac{r - r_+}{r + r_+} \right)
\left( \frac{r + r_-}{r - r_-} \right)^{\ka_+/\ka_-}
\ee
and for $r<r_+$, $r(T^-,X^-)$ is determined by
\be
T^{-2}-X^{-2} = -e^{-2\ka_- r_*} = -e^{-2\ka_- x_0^\prime
  \tan^{-1}(r/r_0)} \left( \frac{r_+ + r}{r_+ - r} \right)^{\ka_-/\ka_+}
\left( \frac{r_--r}{r_-+r} \right).
\ee
However, the boundaries and
singularities are covered by different coordinate patches, which
makes comparison of them impossible.

One can instead follow~\cite{chan:85} and use the coordinates
defined in (\ref{eqn:kruskal}) for the regions $r>r_+$ \emph{and}
$r<r_-$.  In the region $r_- < r < r_+$, we instead define the light
cone coordinates $u=t+r_*,v=-(t-r_*)$ and use \be U = e^{\ka_+ u} =
T-X, \qquad V = e^{\ka_+ v} = T+X. \ee The metric can be written
everywhere as \be \d s^2 = \frac{|f(r)|}{\ka_+^2} e^{-2\ka_+r_*}
(-\d T^2 + \d X^2 ) \ee which is analytic except at the inner
horizon $r=r_-$.  The radial coordinate $r$ is determined implicitly
by
\be
T^2 - X^2 = \mp e^{2\ka_+x_0^\prime
\tan^{-1}(r/r_0)} \left(
  \frac{|r-r_+|}{r+r_+}\right) \left(
  \frac{r+r_-}{|r-r_-|}\right)^{\ka_+/\ka_-}
\ee the $\mp$ sign corresponding to the regions $r>r_+$ and $r<r_-$,
and $r_- < r < r_+$, respectively.

Now we can compare the boundaries and singularities
in the first set of coordinates.  In the limits
\begin{eqnarray}
r &\ra& \infty \qquad \Rightarrow \qquad T^2-X^2
\ra -e^{\pi \ka_+
x_0^\prime}< -1, \\
r &\ra& 0 \qquad \Rightarrow \qquad T^2-X^2 \ra -1
\end{eqnarray}
so in a $(T,X)$ spacetime diagram, the hyperboli
representing the singularities will be further away
from the origin than the hyperboli representing the
boundaries.  This is opposite to the Schwarzschild
case~\cite{stanford}.  Following those arguments,
one sees that on the resulting Penrose diagram if
we draw the singularity $r=0$ as a vertical line,
the boundary $r=\infty$ must be bowed out.
Alternatively, we can use a conformal
transformation to make the singularity bowed out
while keeping the boundary vertical.  The latter
case is shown schematically in figure 2.  (In this
case, to bring the boundaries of the spacetime to a
finite coordinate distance we can, for example, let
$U = e^{\pi \ka_+ x_0^\prime/2} \tan \td{u}$ and $V
= e^{\pi \ka_+ x_0^\prime/2} \tan \td{v}$.)


\sect {Neutral Correlation Functions}

We now discuss features of correlation functions of operators
which are electrically neutral. When we put insertions on both
boundaries, the correlation functions are dominated (in the limit
of high dimensional operators) by spacelike geodesics, as was
explained in \cite{juan}. Such geodesics are only sensitive to the
metric, and not to the background electric field.  In the present
section, we can restrict ourselves to symmetric
geodesics (which reach the boundaries at the same value of
$t_L=t_0$), since all others can be obtained by time translation.

We start in subsection 3.1 by discussing the qualitative features,
before presenting a more quantitative calculation of the correlation
functions.  As in the null case above, the geodesics are specified by
the value of the energy
$E$.  In subsection 3.2, we turn to discussing in detail the relation
between $E$ and the boundary time, $t(E)$, and in subsection 3.3 we compute
the length of the geodesics, which will determine the correlators in
our approximation.


\subsection{Qualitative Features}

\subsubsection*{Radial Spacelike Geodesics}

For radial spacelike geodesics, and normalizing appropriately, the
Lagrangian is \be {\cal L} = -f(r) \dot{t}^2 + \frac{\dot{r}^2}{f(r)} = 1.
\ee The geodesic equations are \be \dot{t} = \frac{E}{f(r)}, \qquad
\dot{r}^2 = E^2 + f(r) \equiv E^2 - V(r), \ee the latter describing
a particle of unit mass and of energy $E^2$ moving in an effective
potential $V(r)=-f(r)$. General properties of the
geodesics\footnote{The uncharged, non-rotating geodesics discussed
here resemble geodesics in the AdS-Schwarzschild geometry which
carry angular momentum~\cite{stanford}.} can thus be read off from
figure 3, in which the effective potential is shown.

\begin{figure}[ht]
         \beginlabels\refpos 0 0 {}
                     \put 150 -5 {V(r)}
                     \put 310 -140 {r}
                     \put 140 -50 {V_{\rm max}}
                     \put 200 -135 {r_-}
                     \put 268 -135 {r_+}
         \endlabels
         \centerline{
         \psfig{figure=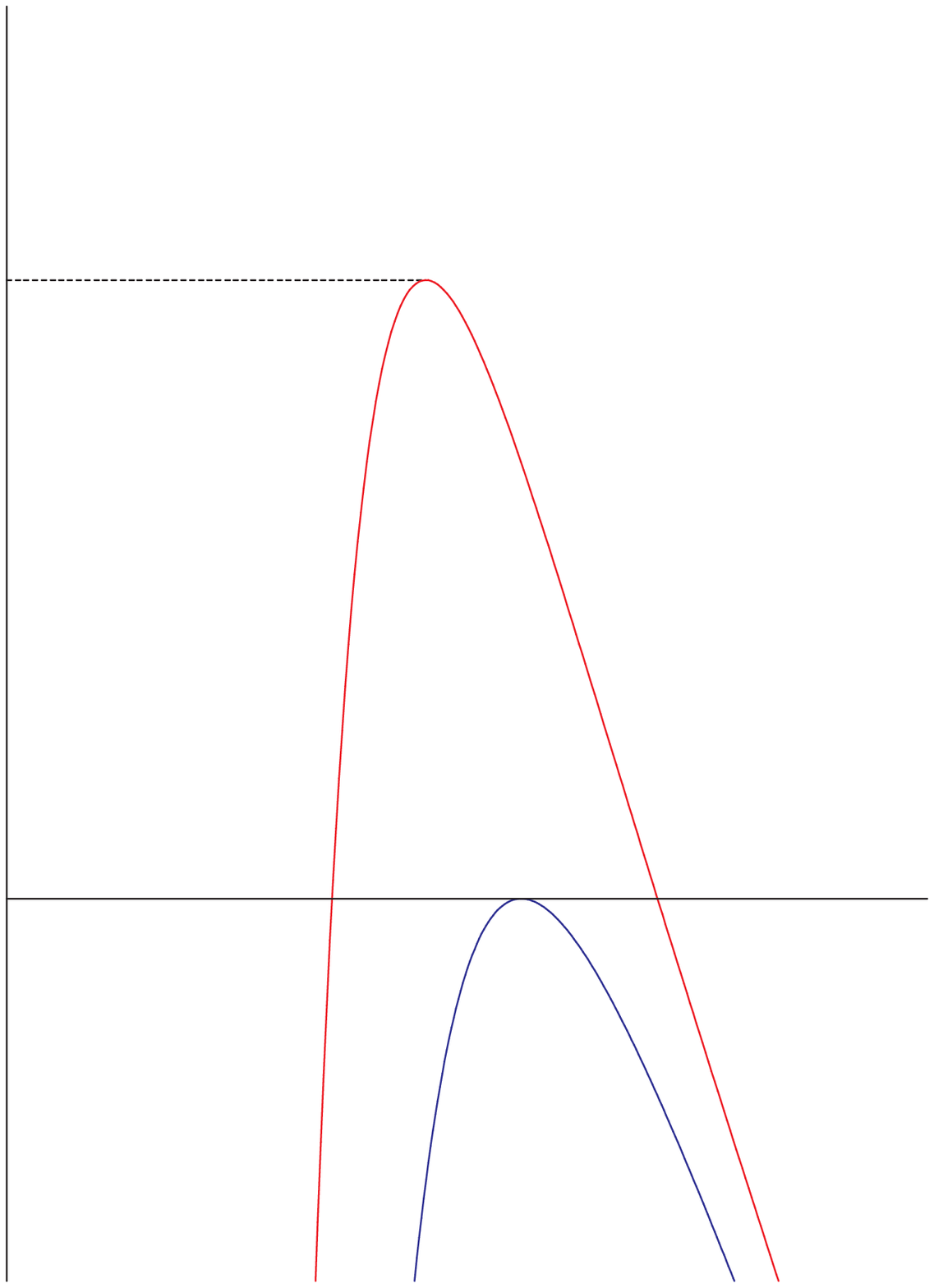,height=2.7in}}
         \vskip0.5cm
         {\footnotesize Figure 3.  The effective potential in the
         non-extreme case is shown in
         red, and the extreme case in blue.}
\end{figure}

The extreme case is relatively uninteresting since there is no
region between the horizon and the singularity which geodesics can
probe.  Geodesics with $E^2>0$ fall all the way into the
singularity; those with $E^2=0$ reach the single event horizon
before returning to the (same) boundary.

The non-extreme case has the following qualitative
features:
\begin{itemize}

\item All geodesics with $E^2 \le V_{\rm max}$ return to the
asymptotic region $r = \infty$, so will connect the two boundaries. The
endpoints lie on the two opposite boundaries, as can be seen by
inspecting $t(\lambda)$.

\item Geodesics with $0 < E^2 \le V_{\rm max}$ will penetrate the
outer horizon, and reach some minimal turning point
between the two horizons.  They cannot penetrate the
inner horizon, and will in fact accumulate at some
finite distance from it (the radius for which
the potential attains its maximum $V_{\rm max}$),
as $E^2 \ra V_{\rm max}$.

\item Geodesics with $E=V_{\rm max}$ penetrate the furthest before
returning to the boundary.  These geodesics will turn out to have
diverging boundary times.

\item Geodesics with $E^2 >V_{\rm max}$ will fall into the
singularity, where our approximation breaks down.   As the
boundary time diverges before we reach this range of $E^2$,  it
seems that such trajectories do not contribute to any gauge
theory process.

\end{itemize}

Apparently, at least for the observables
we are interested in, the
region beyond the Cauchy horizon is not encoded in the gauge
theory.  Similar conclusions about three dimensional rotating
black holes have been obtained in \cite{simon}.  Moreover, the
geodesics do not even reach the Cauchy horizon
itself, reaching at most some minimal distance\footnote{This minimal
distance can be large for far from extremal black holes, and goes to zero
in the extremal limit.} from it. We will see later that, even when
perturbing the spacetime, the observables of interest are screened
from any dramatic behaviour associated with the Cauchy horizon by this
geometrical effect.


\subsubsection*{Rotating Spacelike Geodesics}

Let us add an angular momentum $L$, conjugate to the azimuthal angle
$\p$ on the boundary three-sphere\footnote{For more general
rotation, one replaces $L^2$ by the relevant Casimir operator.}. It
is the conserved quantity associated with the Killing vector
$\del/\del \p$, the relevant geodesic equation being \be \dot{\p} =
\frac{L}{r^2}. \ee In this case, the effective potential is modified
to $V(r) = -f(r) (1-L^2/r^2)$ which, in addition to $r_+$ and $r_-$,
has an additional root at $r=|L|$.  A sketch of one possibility is
given in figure 4.

\begin{figure}[ht]
          \beginlabels\refpos 0 0 {}
                     \put 150 -5 {V}
                     \put 310 -82 {r}
                     \put 140 -63 {V_{\rm max}}
                     \put 190 -77 {r_-}
                     \put 265 -77 {r_+}
                     \put 228 -92 {L}
         \endlabels
         \centerline{
         \psfig{figure=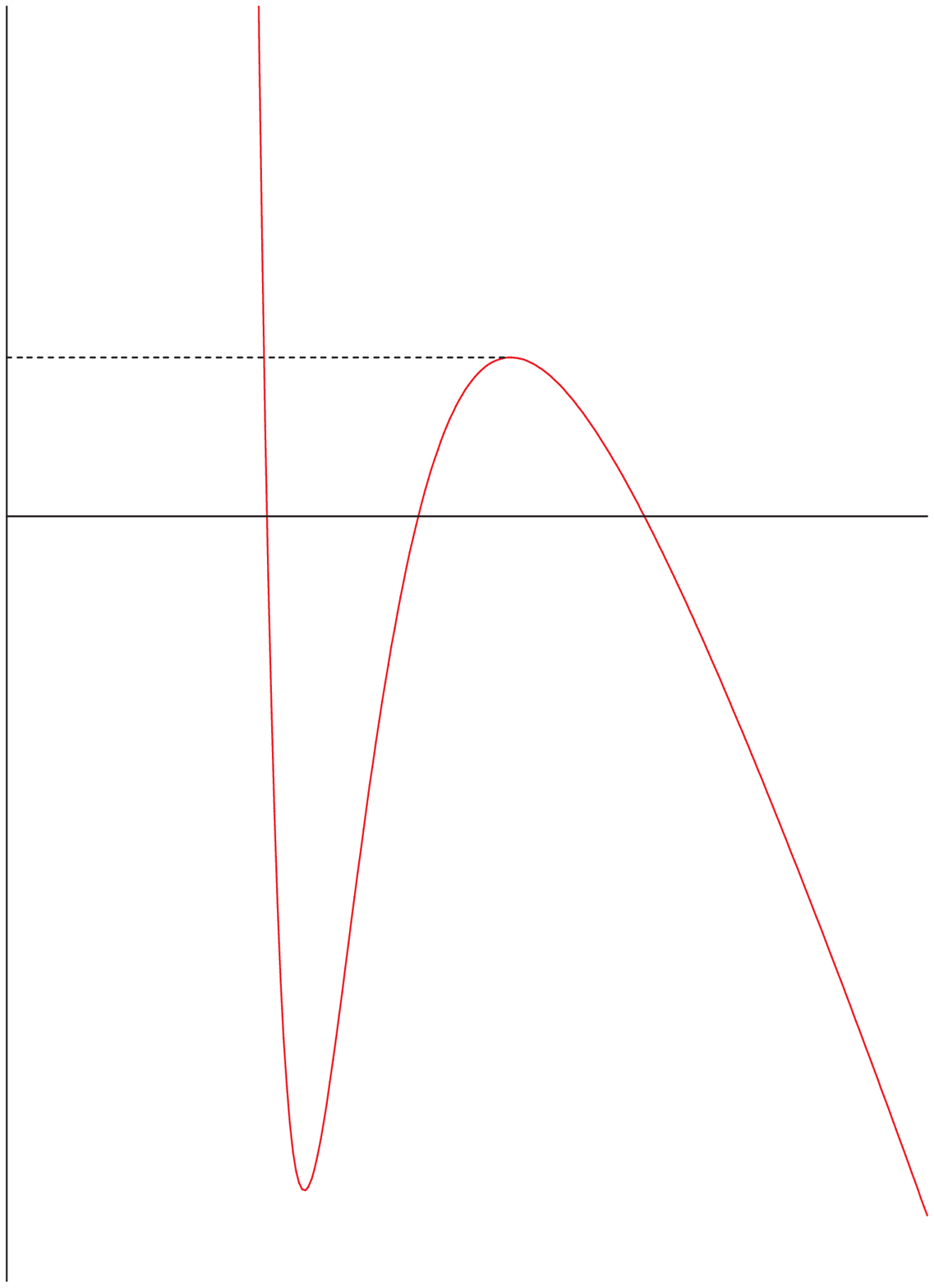,height=2.7in}}
         \vskip0.5cm
         {\footnotesize Figure 4.  The effective potential for a
            rotating geodesic with $x_- < L^2 < x_+$.  Other choices
            of the relative magnitude of $L^2$, $x_-$ and $x_+$ simply
            switch the roots.  If $L^2$ is equal to either $x_-$ or
            $x_+$, then those two roots will coalesce.  The local
            maximum $V_{\rm max}$ is also shown.}
\end{figure}

In the vicinity of the singularity at $r=0$, the
potential is very different to the non-rotating
case: for an arbitrarily small angular momentum, it
turns around at small values of $r$. Consequently,
for sufficiently large $E$ one can come arbitrarily
close to the singularity and still return to the
asymptotic region $r=\infty$; naively the
corresponding two-sided correlators are sensitive
to the region inside the inner horizon.

However, a closer inspection reveals that such geodesics are
irrelevant for the two-sided correlators.  Once again, as
$E^2\rightarrow V_{\rm max}$, the boundary time diverges. In
addition, the behaviour of $t(\lambda)$ shows that these geodesics
connect two boundary points on the {\it same} side of the Penrose
diagram. It is not clear what role, if any, such components play in
the field theory dual, since inner horizon instabilities are likely
to change the causal structure and eliminate these additional
boundaries.

The qualitative details of the rotating
trajectories with $E^2 \le V_{\rm max}$ depend only
on the relative values of $r_+$ and $L$.  For $|L|
> r_+$ the geodesics do not penetrate the outer
horizon, making them uninteresting for our
purposes.  On the other hand, for $|L| < r_+$ the
situation is basically unchanged from the
non-rotating case, the geodesics for which $0<E^2
\le V_{\rm max}$ being relevant to us.


\subsection{Boundary Time}

In this subsection we describe the relation between
(the real part of) the boundary time $t_0$ and the
energy $E$.  We are interested in symmetric
geodesics which start at a point $t_0$ on the
boundary, and end at the point $-(t_0 + i \beta/2)$
on the other boundary.  Taking $r_E$ to denote the
turning point for a trajectory of energy $E^2$, we
have \be E^2 + f(r_E) = 0. \label{eqn:tp} \ee As
in~\cite{stanford}, we must have $t_L(r_E)=0$ and
$t_E(r_E)=-\beta/4$.  Such geodesics satisfy \be
\label{relation1} t_0 + i \frac{\beta}{4} = E
\int_{r_E}^\infty \frac{\d r}{V(r) \sqrt{E^2 -
V(r)}} \ee where $V(r)$ is the effective potential
described above, for either rotating or
non-rotating geodesics. Since the qualitative
features are similar, we will only be concerned
with the latter, for which $V(r)=-f(r)$.

The two features prominent in the analogous discussion of
\cite{stanford}, for the AdS-Schwarzschild metric, are the branch cut
at $t_0=0$ (corresponding to $E=0$) and the existence of a
critical value $t_0 =-t_c$ (corresponding to $E=\infty$), where
the correlators naively diverge. The latter feature was
interpreted as a signal of the singularity encoded in gauge theory
correlation functions, the former making that encoding a subtle
one.  In our discussion these features are no longer present.  The
special values of the energy are $E=0$ and $E^2=V_{\rm max}$, and
we discuss them both below.

For general energies, one can compute (\ref{relation1}) in terms
of elliptic integrals\footnote{We use the notation
of~\cite{elliptic} for the elliptic integrals.}, which we do in
the appendix.  The result will depend both on the
roots of the cubic $\Delta(x)$ and on the roots of the other cubic
appearing in the denominator of (\ref{relation1}),
\be
\td{\Delta}(x) = E^2 x^2 + (x+x_0)(x-x_-)(x-x_+) \equiv
(x+\td{x}_0)(x-\td{x}_-)(x-\td{x}_+)
\label{eqn:newcubic}
\ee
where the roots $\td{x}_+ > \td{x}_- > 0 > -\td{x}_0$ and $\td{x}_+=r_E^2$.

From the appendix, and dropping the pure imaginary
piece, we have\footnote{This expression is actually valid for
\emph{any} geodesic, not just symmetric ones.  We simply have to replace $t_0$
with $t_{\rm b} - t_{\rm tp}$, these values denoting, respectively, the
values of $t_L$ at the boundary and at the turning point.}
\[
t_0 = -E \frac{1}{ \sqrt{ \td{x}_+ (\td{x}_- +
 \td{x}_0) } } \left[
 \frac{x_0^3(\td{x}_+-\td{x}_-)}{(\td{x}_++x_0)(\td{x}_-+x_0)(x_0+x_-)(x_0+x_+)}~\Pi(\td{\p},\td{\alpha}_0^2,\td{k})
 \right.
\]
\[
+\frac{x_-^3(\td{x}_+-\td{x}_-)}{(\td{x}_+-x_-)(\td{x}_--x_-)(x_0+x_-)(x_+-x_-)}
~\Pi(\td{\p},\td{\alpha}_-^2,\td{k})
\]
\[
-\frac{x_+^3(\td{x}_+-\td{x}_-)}{(\td{x}_+-x_+)(\td{x}_--x_+)(x_0+x_+)(x_+-x_-)}
~\Pi(\td{\p},\td{\alpha}_+^2,\td{k})
\]
\be
\left. +
\frac{\td{x}_-^3}{(\td{x}_-+x_0)(\td{x}_--x_-)(\td{x}_--x_+)}
 ~{\rm F}(\td{\p},\td{k}) \right]
\label{eqn:time}
\ee
where
\[
\td{\alpha}_0^2 = \left(
\frac{\td{x}_++\td{x}_0}{\td{x}_-+\td{x}_0} \right)\left(
\frac{\td{x}_-+x_0}{\td{x}_++x_0} \right), \qquad \td{\alpha}_-^2 =
\left( \frac{\td{x}_++\td{x}_0}{\td{x}_-+\td{x}_0} \right)\left(
\frac{\td{x}_--x_-}{\td{x}_+-x_-} \right),
\]
\be
\td{\alpha}_+^2 = \left( \frac{\td{x}_++\td{x}_0}{\td{x}_-+\td{x}_0}
\right)\left( \frac{\td{x}_--x_+}{\td{x}_+-x_+} \right), \qquad
\td{k} = \sqrt{\frac{\td{x}_-}{\td{x}_+}
\left(\frac{\td{x}_++\td{x}_0}{\td{x}_-+\td{x}_0} \right)}, \qquad
\td{\p} =
\sin^{-1}\sqrt{\frac{\td{x}_-+\td{x}_0}{\td{x}_++\td{x}_0}}.
\ee

This is a fairly unilluminating expression, but more information can
be seen if we plot $t(E)$ numerically, for
various choices of black hole parameters. We find two distinct
types of behaviour, shown in figure 5:

\begin{itemize}

\item The left-hand plot is representative of the behaviour for
$x_+ \sim x_-$, where the  black hole is near-extremal. The
boundary time $t$ is then a monotonically increasing function of
energy $E$.

\item The right-hand plot, on the other hand, represents the
behaviour for $x_+ >> x_-$, this limit being one of large mass and
small charge.

\item In both cases, we clearly see a divergence in
$t$ as $E^2 \ra V_{\rm max}$.

\end{itemize}

\vskip 0.5cm\begin{figure}[ht]
         \beginlabels\refpos 0 0 {}
                     \put 40 -3 {t}
                     \put 288 -3 {t}
                     \put 180 -182 {E}
                     \put 428 -125 {E}
                     \put 150 -3 {\sqrt{V_{\rm max}}}
                     \put 400 -3 {\sqrt{V_{\rm max}}}
         \endlabels
         \centerline{
         \psfig{figure=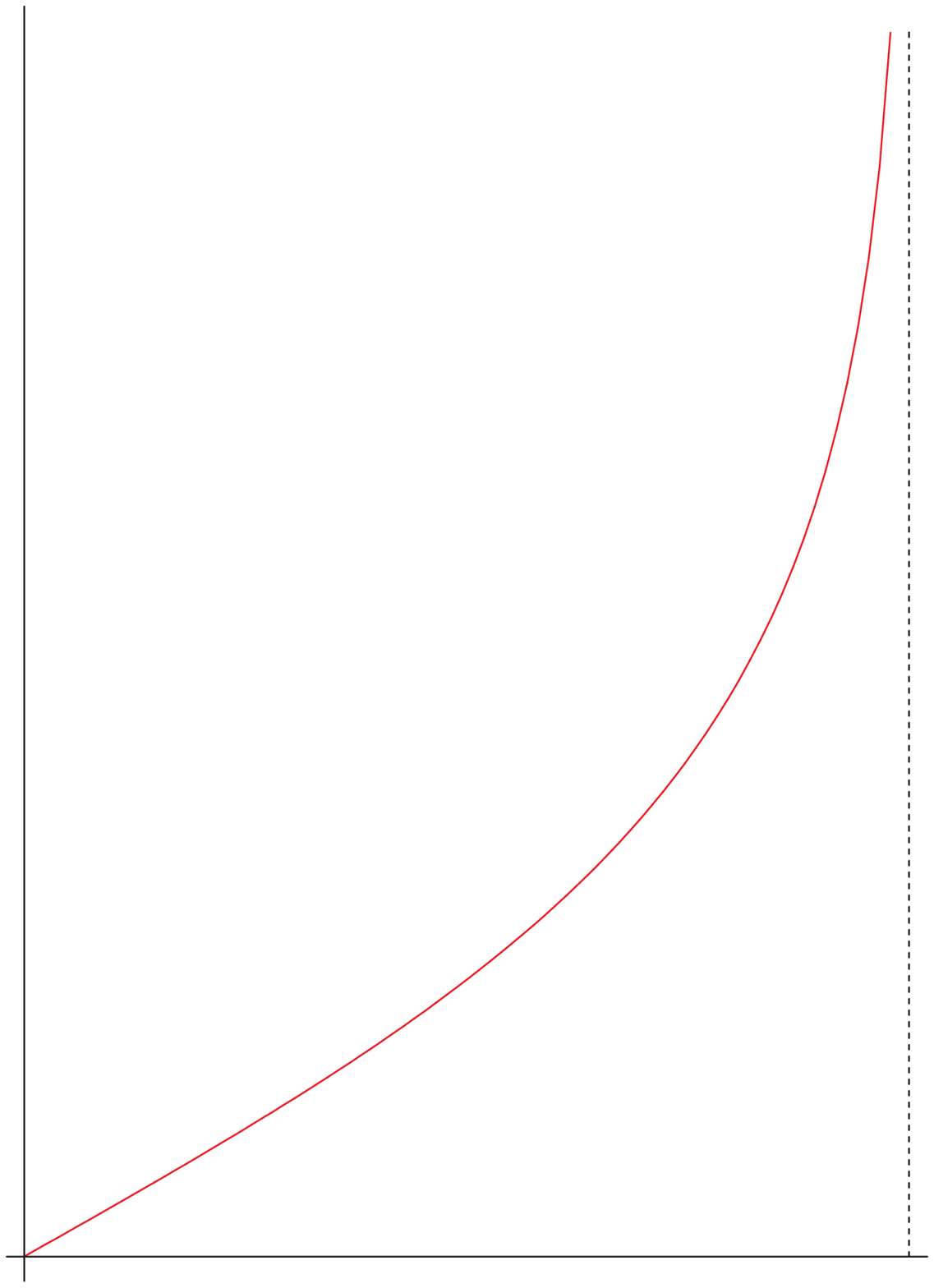,height=2.5in}
         \hskip4.0cm
         \psfig{figure=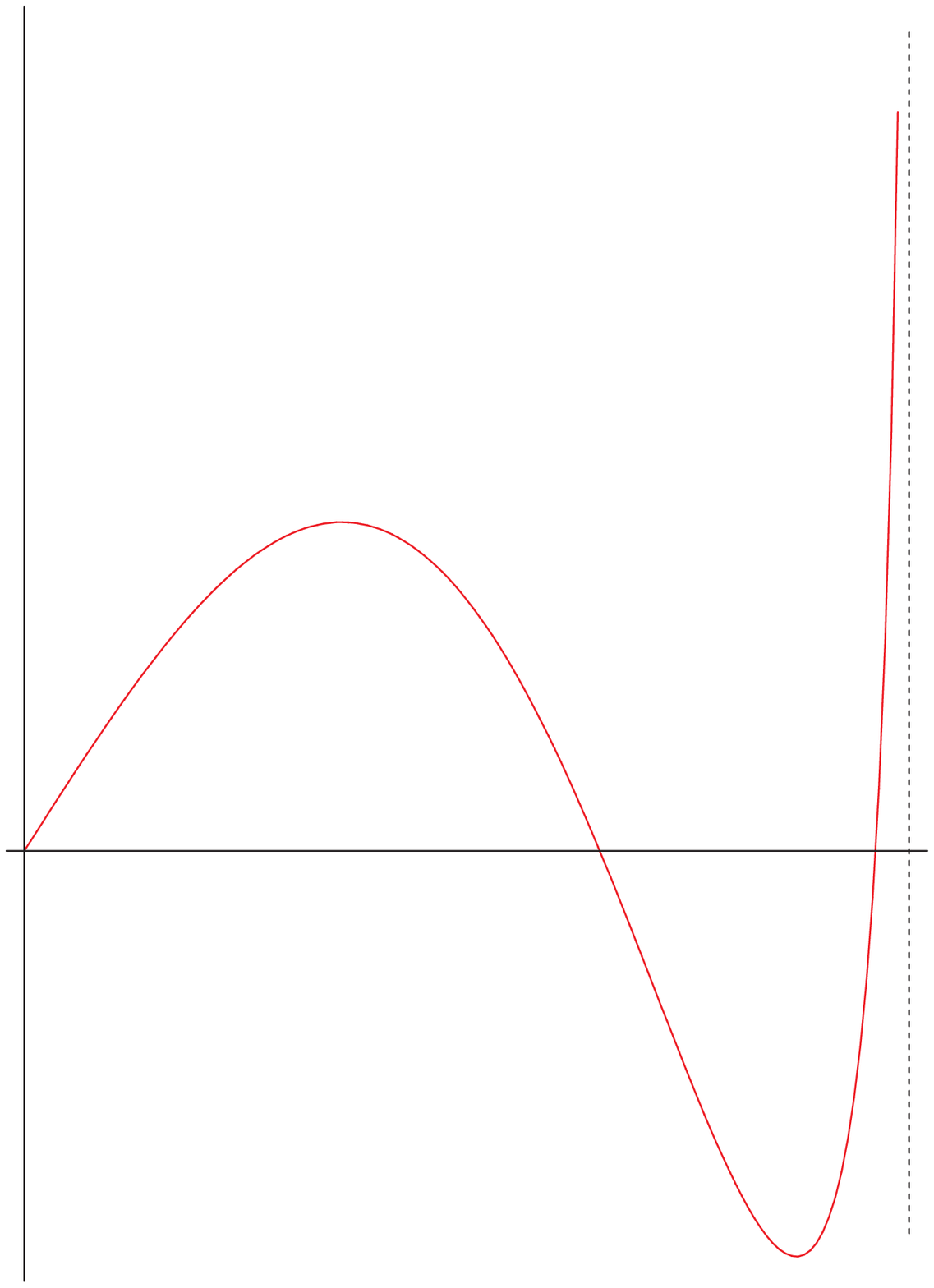,height=2.5in} }
         \vskip0.5cm
         {\footnotesize Figure 5.  The left-hand plot shows $t(E)$ for
                     $x_+ \sim x_-$ whereas the right-hand plot has
                     $x_+ >> x_-$.  The dotted lines indicate $E^2 =
                     V_{\rm max}$.}
\end{figure}

There is thus a range of parameter space in which, for each value
of boundary time $t$, there exists a unique geodesic connecting
the two boundary points (the left-hand plot in figure 5). However,
outside of this range (the right-hand plot in figure 5), there are
values of boundary time for which there are multiple geodesics
connecting the two boundaries, each with a different value of $E$.
Which one dominates the path integral will depend on the relative
proper lengths.

Since the cross-over between different saddle points will give rise to
non-analyticity of the correlation function,
we see that the existence or otherwise of non-analytic
behaviour depends on the black hole parameters, $x_\pm$, in an interesting
way\footnote{Similar changes of behaviour as a function of
parameters was noticed and interpreted in \cite{RN}, and it would be
interesting to relate these two types of phase transition.}. We comment
further on this feature below (in subsection (3.3)), when discussing
the proper length of these spacelike geodesics.

The precise point of cross-over between these two types of
behaviour, as a function of parameters, is difficult to pin down
analytically but is possible to analyse numerically. For example,
figure 6 shows a region of $x_\pm$ parameter space, the circles
denoting that $t$ is a monotonically increasing function of $E$ for
those choices of parameters, and the crosses denoting that $t$ has a
turning point\footnote{ It is curious that the boundary between
these two types of behaviour is a straight line.}.

\vskip 0.5cm\begin{figure}[ht]
         \beginlabels\refpos 0 0 {}
                     \put 158 3 {x_+}
                     \put 320 -215 {x_-}
         \endlabels
         \centerline{
         \psfig{figure=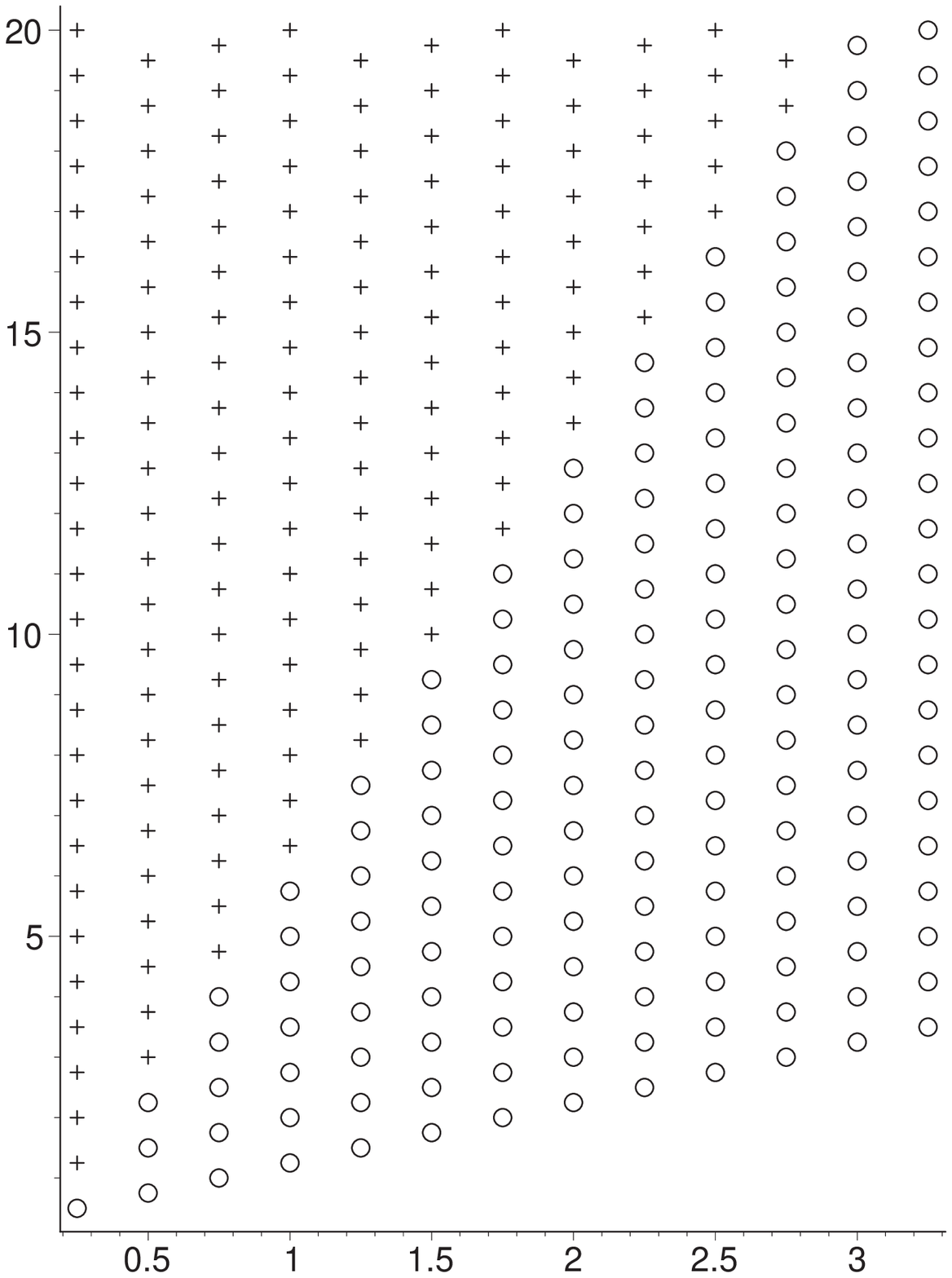,height=3.0in}}
         {\footnotesize Figure 6.  This plot shows how $t(E)$ behaves as a
         function of the parameters $x_\pm$.  Circles denote points
         for which $t$ is a monotonically increasing function of $E$,
         and crosses denote points for which $t$ has a turning
         point for some value of $E$.}
\end{figure}


\subsubsection*{Small $E$ Geodesics}

It is easy to see that $E=0$ corresponds to
$t_0=0$, both for rotating and non-rotating
geodesics. For $E=0$, we have
\be
\dot{t} = 0 \qquad \Rightarrow \qquad
 t(\la)=0, \qquad \dot{r}^2 = f(r),
\ee so these geodesics start at the boundary, pass straight through
the Penrose diagram and end at the other boundary, without ever
penetrating the horizon.

To derive the small $E$ behaviour of (\ref{eqn:time}), we need the
$E$-dependence of the roots $\td{x}_0, \td{x}_-$ and $\td{x}_+$.
This is determined by the following set of equations:
\begin{eqnarray}
&& \td{x}_0 - \td{x}_- - \td{x}_+ = x_0 - x_- - x_+ + E^2,\nonumber \\
&& \td{x}_- \td{x}_+ - \td{x}_0 \td{x}_+ - \td{x}_0
\td{x}_- = x_- x_+
- x_0 x_+ - x_0 x_-, \\
&& \td{x}_0 \td{x}_- \td{x}_+ = x_0 x_- x_+. \nonumber
\end{eqnarray}
Since we are interested only in small $E$, we can
solve these by taking
\begin{eqnarray}
\td{x}_0 &=& x_0 + (1 + a_- + a_+) E^2 + {\cal O} (E^4), \nonumber \\
\td{x}_- &=& x_- + a_- E^2 + {\cal O} (E^4), \label{eqn:E-roots} \\
\td{x}_+ &=& x_+ + a_+ E^2 + {\cal O} (E^4) \nonumber
\end{eqnarray}
where \be a_\pm = \mp \frac{r_\pm}{\ka_\pm}. \ee It is important
that the possible order $E$ terms in (\ref{eqn:E-roots}) necessarily
vanish, as long as $x_+ \ne x_-$, which in turn implies that all odd
powers of $E$ also vanish.  We then have \be \td{\p} = \p + \p_1 E^2
+ {\cal O}(E^4), \qquad \td{k} = k + k_1 E^2 + {\cal O}(E^4), \qquad
\td{\alpha}^2 = \alpha^2 + \alpha_1^2 E^2 + {\cal O} (E^4) \ee (for each
of the $\td{\alpha}$s), and where the quantities without a tilde are
independent of $E$, being functions of the roots $x_0, \, x_-$ and
$x_+$ only. The elliptic integrals can then be Taylor expanded in
each of their arguments using the well-known expressions for the
various derivatives~\cite{elliptic}.

The net result is that \be t_0 = t_1 E + {\cal O}(E^3)
\label{eqn:smallEt} \ee for some non-vanishing constant $t_1$.
This expression for $t(E)$  is similar to the finite mass black
hole of~\cite{stanford}, and in particular is an analytic function
near $E=0$.


\subsubsection*{The Regime $E^2 \rightarrow V_{\rm max}$}

The integrand in (\ref{relation1}) is divergent at the turning point
of the trajectory, where $E^2=V(r)$.  This singularity is integrable
for $E^2 \neq V_{\rm max}$, leading to finite boundary time $t_0$
for $0< E^2 < V_{\rm max}$. On the other hand, when $E^2 =V_{\rm
max}$, the two positive roots of $\td{\Delta}(x)$ coincide
($\td{x}_+=\td{x}_-$), leading to a logarithmic divergence in
(\ref{relation1}).    We emphasise that this is true for both the
rotating and non-rotating geodesics.

The divergence is shown clearly in figure 5, but we can see it
analytically from the exact result (\ref{eqn:time}) in the following
way.  Take $\td{x}_- = \td{x}_+ - \vep$ and note that $\td{k}$ and
each $\td{\alpha}^2$ go to one faster than $\td{\p}$ goes to
$\pi/2$, since \be \td{k} = 1 + {\cal O}(\vep), \qquad \td{\alpha}^2 = 1 +
{\cal O}(\vep), \qquad \td{\p} = \frac{\pi}{2} + {\cal O}(\vep^{1/2}). \ee Then
note that \be \td{x}_+ - \td{x}_- =
\frac{\td{x}_+(\td{x}_-+\td{x}_0)}{\td{x}_0} (1-\td{k}^2) \ee and
use~\cite{elliptic}
\be
\Pi(\phi,1,k) = \frac{1}{(1-k^2)} \left( (1-k^2) \,
F(\p,k) - E(\p,k)
  + \tan \p \sqrt{1-k^2 \sin^2 \p} \right).
\ee The problematic $(1-k^2)$ factors cancel, giving \ba t_0 &=&
-E\frac{\td{x}^{5/2}}{(\td{x}_++\td{x}_0)^{3/2}(\td{x}_+-x_-)(\td{x}_+-x_+)}
\, \ln (\tan \td{\p} + \sec \td{\p} ) \nonumber \\ &\sim& - \ln
(\cos\pi/2) + {\rm finite}, \ea since $x_- < \td{x}_+ < x_+$.

Since the boundary time diverges as $E^2 \rightarrow V_{\rm max}$, the
gauge theory does not seem to encode the regime $E^2>V_{\rm max}$.


\subsection{Correlation Functions}

The proper length of both the rotating and
non-rotating geodesics is given by \be L = 2
\int_{r_E}^{r_{\rm max}} \frac{\d r}{\sqrt{E^2 -
V(r)}} \label{eqn:proper} \ee with $V(r)$ being the
relevant potential in each case.  $r_E$ is the same
turning point as in the previous subsection and,
since we will only be concerned with the
non-rotating case, it is again given by solving
(\ref{eqn:tp}).  The upper limit, $r_{\rm max}$, is
a long-distance radial cutoff, dual to the UV
cutoff in the gauge theory.  As we take $r_{\rm
max} \ra \infty$, the integral diverges
logarithmically. To regularise it we subtract the
divergent piece arising in the pure AdS case
(obtained by setting $x_0=1$ and $x_-=x_+=0$ in the
above expression).  This standard process is dual
to renormalisation of the boundary theory
\cite{vijay:99}.


\subsubsection*{$E=0$ Geodesics}

We start with the $E=0$ symmetric geodesics, which approximate the
boundary correlators with $t_0=0$. The proper length of such a
geodesic is \be L = 2 \int_{r_+}^{r_{\rm max}} \frac{\d
r}{\sqrt{f(r)}} = \int_{r_+^2}^{r_{\rm max}^2}  \d x
\sqrt{\frac{x}{(x+x_0)(x-x_-)(x-x_+)} } \label{eqn:length} \ee where
the turning point for $E=0$ is $r=r_+$, the location of the outer
horizon.  We can compute (\ref{eqn:length}) in terms of elliptic
integrals, giving~\cite{elliptic} \be L =
\frac{2}{\sqrt{x_+(x_-+x_0)}}\left[ x_- ~{\rm F}(\phi,k) + (x_+-x_-)
~\Pi (\phi, x_+ k^2/x_-,k ) \right] \ee where \be \phi = \sin^{-1}
\sqrt{ \frac{ (x_-+x_0) (r^2_{{\rm max}} - x_+ )}{
(x_++x_0)(r^2_{{\rm max}} - x_- ) }}, \qquad k = \sqrt{ \frac{ x_-
(x_++x_0) }{ x_+ (x_-+x_0) } } = \sqrt{\frac{\ka_+}{\ka_-}
\frac{r_+}{r_-}}. \ee In the limit $r_{\rm max} \ra \infty$, the
logarithmic divergence appears in $\Pi ( \phi,x_+k^2/x_-,k )$
(whereas ${\rm F}(\phi,k)$ remains finite).  To extract this
divergence, we note that $\alpha^2=x_+ k^2/x_- > 1$, so we can
use~\cite{ab} \be \Pi (\p, \al^2, k) = - \Pi (\p, k^2/\al^2,k ) +
{\rm F}(\p, k) + \frac{1}{2p_1} \ln \left(
\frac{\Delta(\p)+p_1\tan\p}{\Delta(\p)-p_1\tan\p} \right) \ee where
$p_1=\sqrt{(\al^2-1)(1-k^2/\al^2)}$ and $\Delta(\p) =
\sqrt{1-k^2\sin^2\p}$.  Expanding to first order in $1/r_{\rm
max}^2$ gives \be L = \frac{2}{\sqrt{x_+(x_-+x_0)}} \left[ x_+ ~{\rm
F} (\p,k) - (x_+-x_-)~ \Pi ( \p, x_-/x_+, k ) \right] + \ln \left(
\frac{4}{x_0+x_-+x_+} \right) + \ln (r^2_{\rm max}), \ee the
logarithmic divergence being manifest.  The renormalised result is
thus \be L_{\rm ren} = \frac{2}{\sqrt{x_+(x_-+x_0)}} \left[ x_+
~{\rm F} (\p,k) - (x_+-x_-)~ \Pi ( \p, x_-/x_+, k ) \right] + \ln
\left( \frac{4}{x_0+x_-+x_+} \right) \label{eqn:ren} \ee where now
\be \p = \sin^{-1} \sqrt{ \frac{ x_-+x_0 }{ x_++x_0 }}. \ee


\subsubsection*{$0<E< V_{\rm max}$ Geodesics}

As in~\cite{stanford}, as we increase the energy, the geodesic will
penetrate some distance inside the horizon.  The proper length of
the (non-rotating) geodesic is now: \be {\cal L} = 2 \int_{r_E}^{r_{\rm
max}} \frac{\d r}{\sqrt{E^2+f(r)}} = \int_{r_E^2}^{r_{\rm max}^2} \d
x \sqrt{\frac{x}{E^2x^2+(x+x_0)(x-x_-)(x-x_+)}} \label{eqn:Lint} \ee
which will depend on the roots of the cubic $\td{\Delta}(x)$,
defined in (\ref{eqn:newcubic}).  Formally, the result is as in
(\ref{eqn:ren}), but with the $E$-dependent roots: \be L_{\rm ren} =
\frac{2}{\sqrt{\td{x}_+(\td{x}_-+\td{x}_0)}} \left[ \td{x}_+ ~{\rm
F} (\td{\p},\td{k}) - (\td{x}_+-\td{x}_-)~ \Pi ( \td{\p},
\td{x}_-/\td{x}_+, \td{k} ) \right] + \ln \left(
\frac{4}{\td{x}_0+\td{x}_-+\td{x}_+} \right) \ee where \be \td{\p} =
\sin^{-1} \sqrt{ \frac{ \td{x}_-+\td{x}_0 }{ \td{x}_++\td{x}_0 }},
\qquad \td{k} = \sqrt{ \frac{ \td{x}_- (\td{x}_++\td{x}_0) }{
\td{x}_+ (\td{x}_-+\td{x}_0) } }. \ee

To take the small $E$ limit of this expression, we proceed as in the
previous subsection, using the expressions (\ref{eqn:E-roots}). The
upshot is that \be L(E) - L(E=0) = L_1 E^2+{\cal O}(E^3) \ee for some
non-vanishing constant $L_1$.  Using the previous expression
(\ref{eqn:smallEt}) for the small $E$ behaviour of the boundary
time, we see that \be L(E) - L(E=0) \sim t^2. \ee The result is
therefore analytic around $E=0$, the branch cut structure found
in~\cite{stanford} for the infinitely massive black hole being
absent here.  It is more similar to the finite mass black
hole~\cite{stanford}.

On the other hand, as for the computation of the boundary time above,
the integral (\ref{eqn:proper}) will have an
additional logarithmic divergence at the lower
limit for $E = V_{\rm max}$, which we can again extract analytically,
or observe numerically.  In figure 7, we plot $L(E)$, the proper
length as a function of energy, for the two different regimes of the
previous subsection.  The parameters are the same as those in figure
5, the left-hand plot being representative of the behaviour for $x_+
\sim x_-$, and the right-hand plot being representative of the
behaviour for $x_+ >> x_-$.  The divergence as $E^2 \ra V_{\rm max}$
can clearly be seen.

\vskip 0.5cm\begin{figure}[ht1]
         \beginlabels\refpos 0 0 {}
                     \put 30 -3 {L - L(0)}
                     \put 278 -3 {L - L(0)}
                     \put 180 -182 {E}
                     \put 428 -73 {E}
                     \put 150 -3 {\sqrt{V_{\rm max}}}
                     \put 400 -3 {\sqrt{V_{\rm max}}}
         \endlabels
         \centerline{
         \psfig{figure=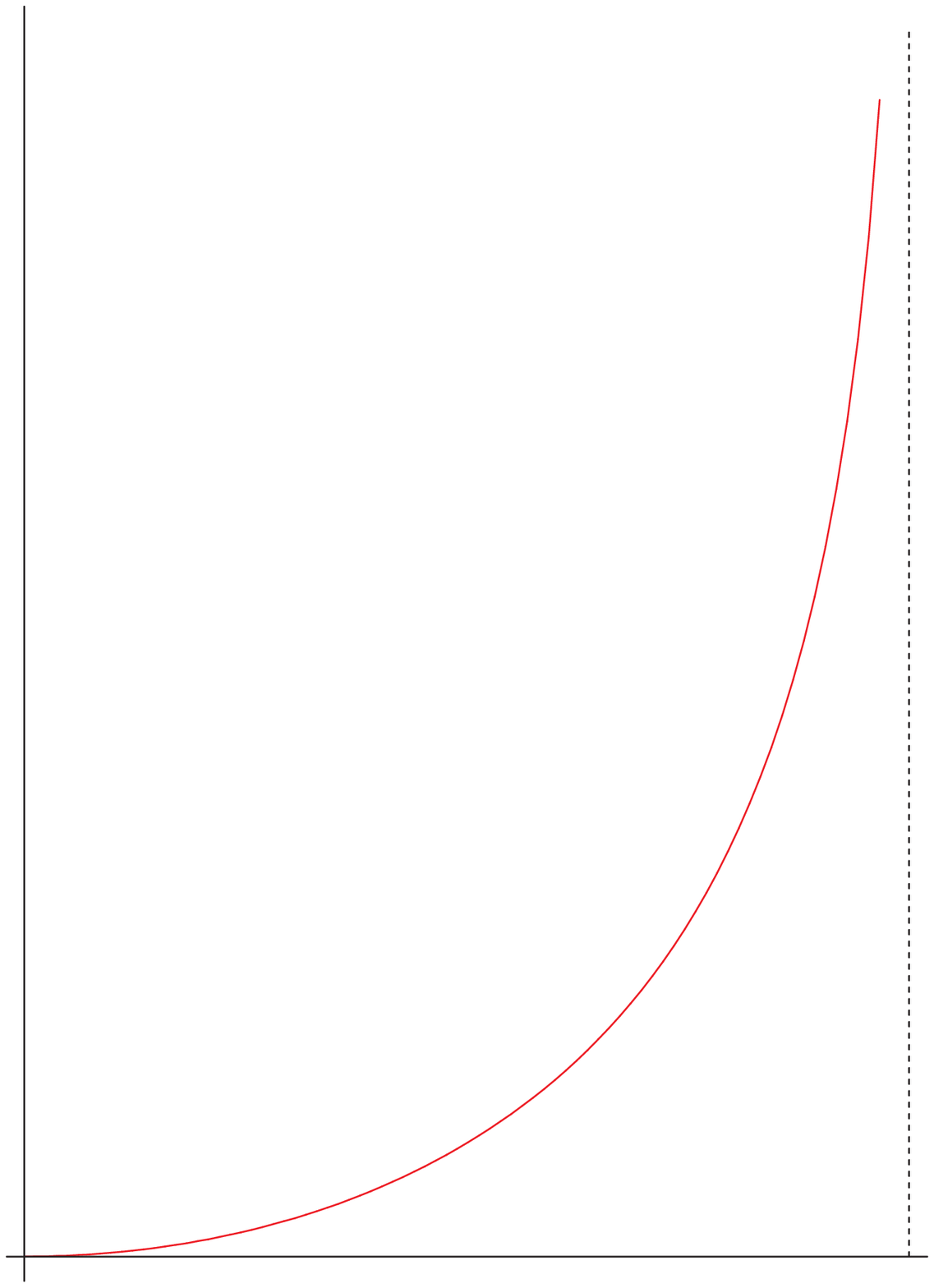,height=2.5in}
         \hskip4.0cm
         \psfig{figure=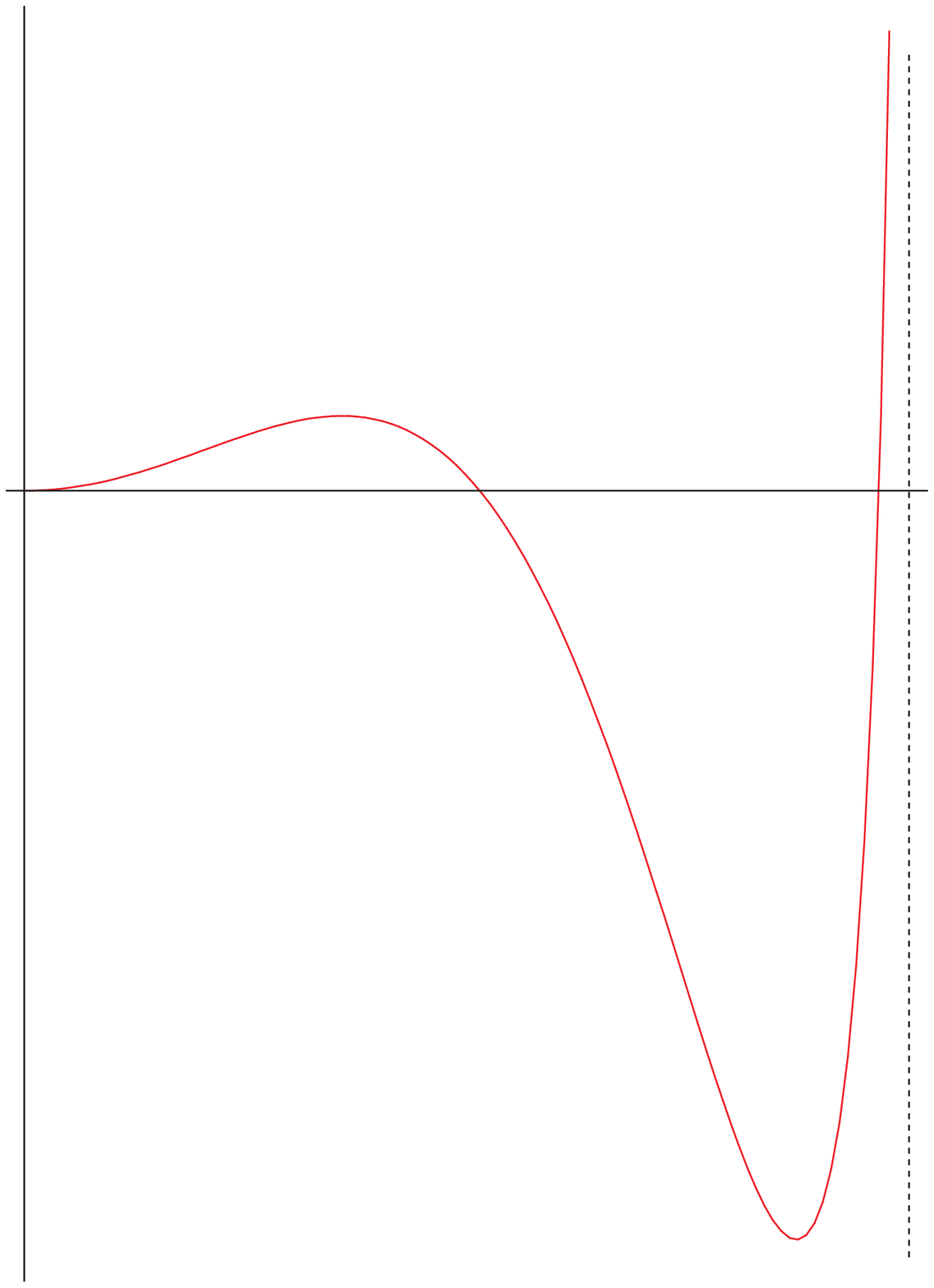,height=2.5in} }
         \vskip0.5cm
         {\footnotesize Figure 7.  The left-hand plot shows
         $L(E)-L(E=0)$ for $x_+ \sim x_-$ whereas the right-hand plot has
         $x_+ >> x_-$.  The dotted lines indicate $E^2 = V_{\rm max}$.}
\end{figure}


\subsubsection*{Non-Analytic Behaviour}

We have found that the subtle features of the infinitely massive
AdS-Schwarzschild black hole are absent in our calculation. This is not
surprising, since even very far from extremality we are expected to
reproduce only the features of the finite mass black hole.

On the other hand we find a new source of non-analytic behaviour which is
much less subtle, coming from multiple {\it real} geodesics connecting
the same boundary points. The precise type of non-analyticity depends
on the parameters of the black hole. The possibilities are shown in
figure 5, the plots of $t(E)$.

For large enough boundary time, there is always a unique geodesic
connecting the boundary points.  As we decrease the value of the
boundary time, there is a critical time $t_{\rm crit}$ (and
corresponding energy $E_{\rm crit}$) for which additional geodesics
appear. This is seen in the right-hand plot of figure 5, for the
case of a black hole far enough from extremality. The type of
non-analyticity then depends on which geodesic has the least action,
which can be read from figure 7, the plots of $L(E)$.  To make this
process more transparent, in figure 8 we plot $L(t)$, the value of
the correlator as a function of boundary time for all branches, and
for different values of the parameters $x_+$ and $x_-$.

\vskip 0.5cm\begin{figure}[ht]
         \beginlabels\refpos 0 0 {}
                     \put 55 0 {L - L(0)}
                     \put 278 0 {L - L(0)}
                     \put 180 -145 {t}
                     \put 400 -140 {t}
                     \put 88 -175 {L - L(0)}
                     \put 358 -175 {L - L(0)}
                     \put 180 -272 {t}
                     \put 400 -205 {t}
         \endlabels
         \centerline{
         \psfig{figure=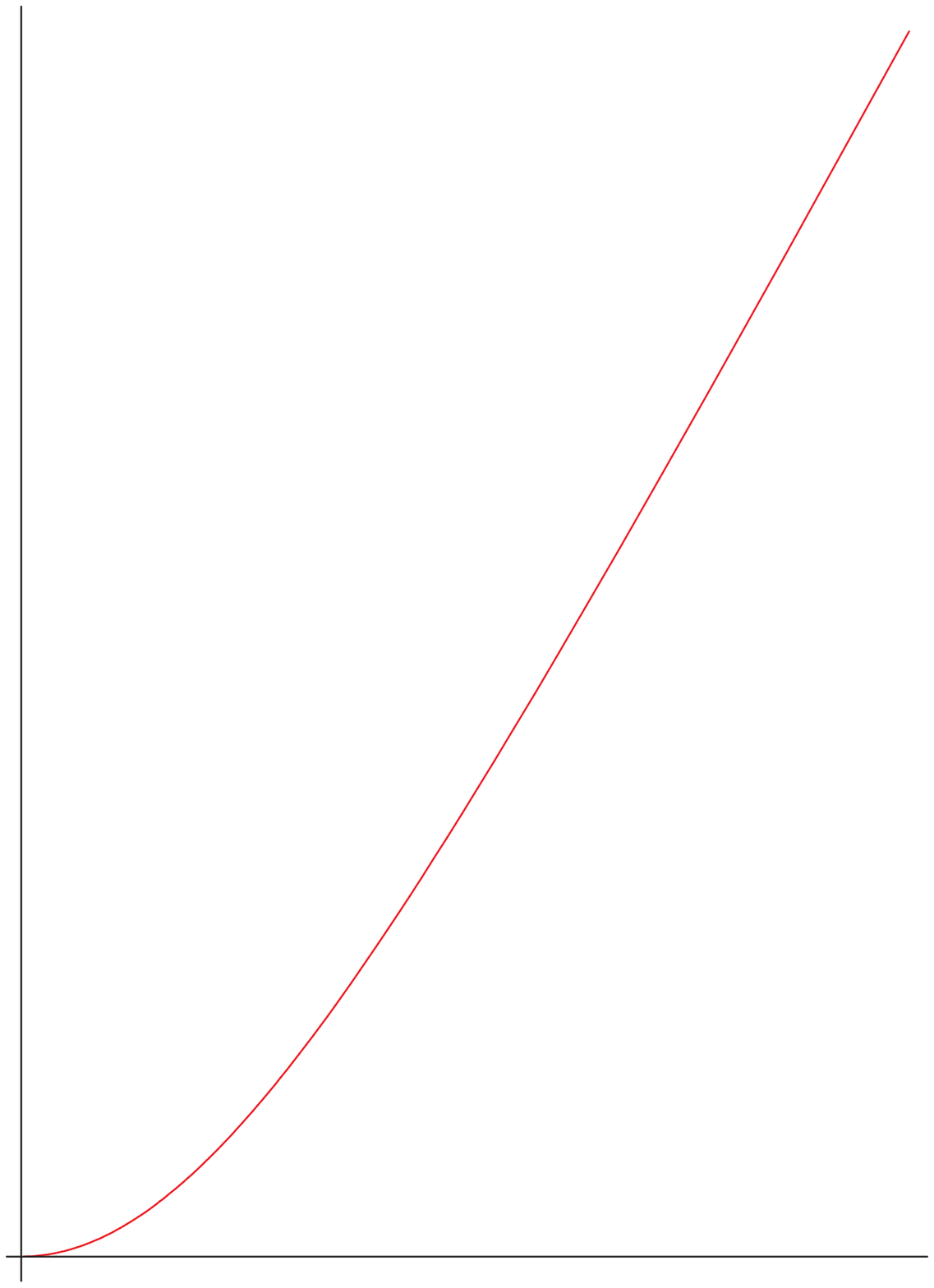,height=2in}
         \hskip4.0cm
         \psfig{figure=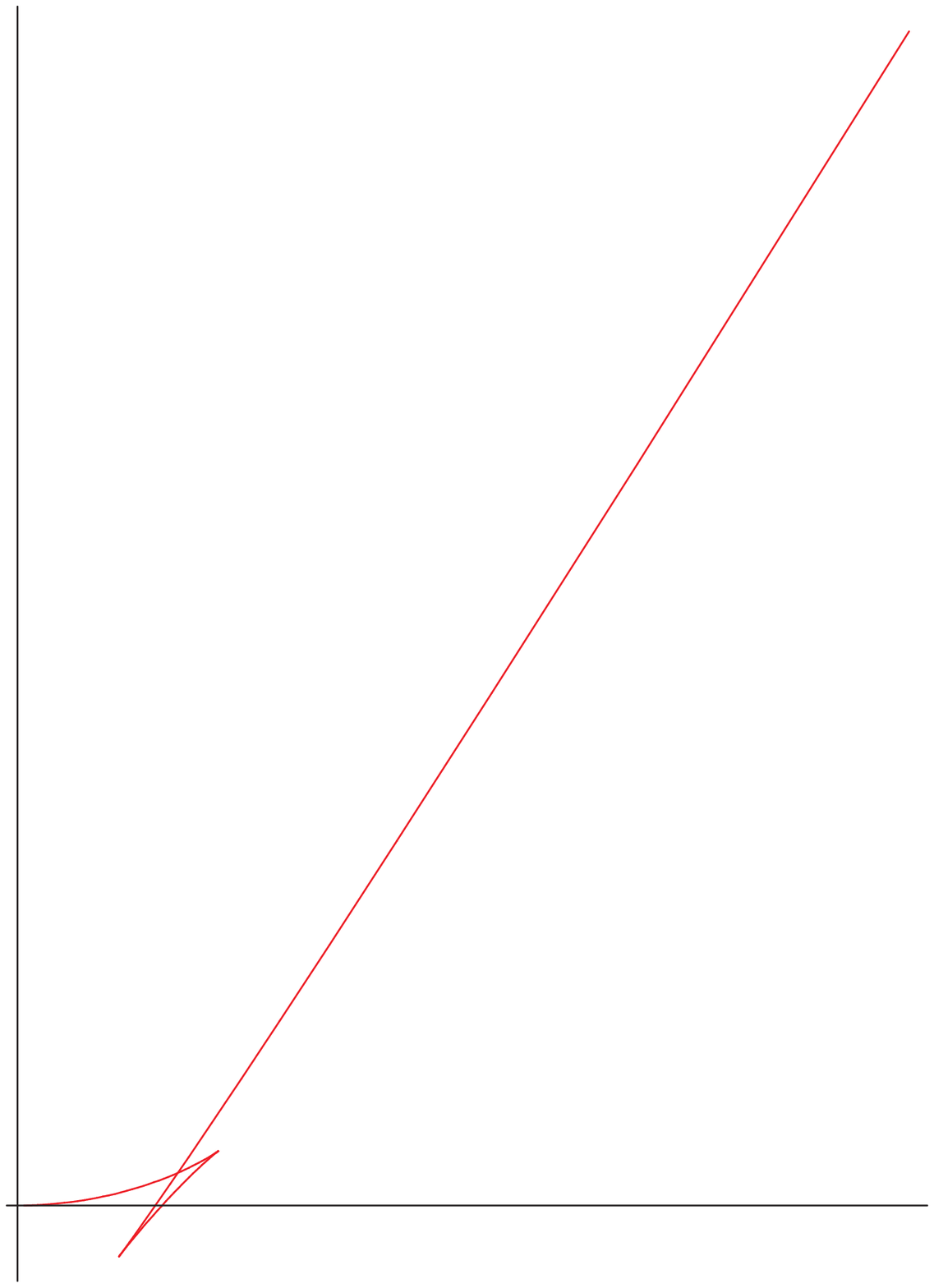,height=2in}  }
         \vskip1.0cm
         \centerline{
         \psfig{figure=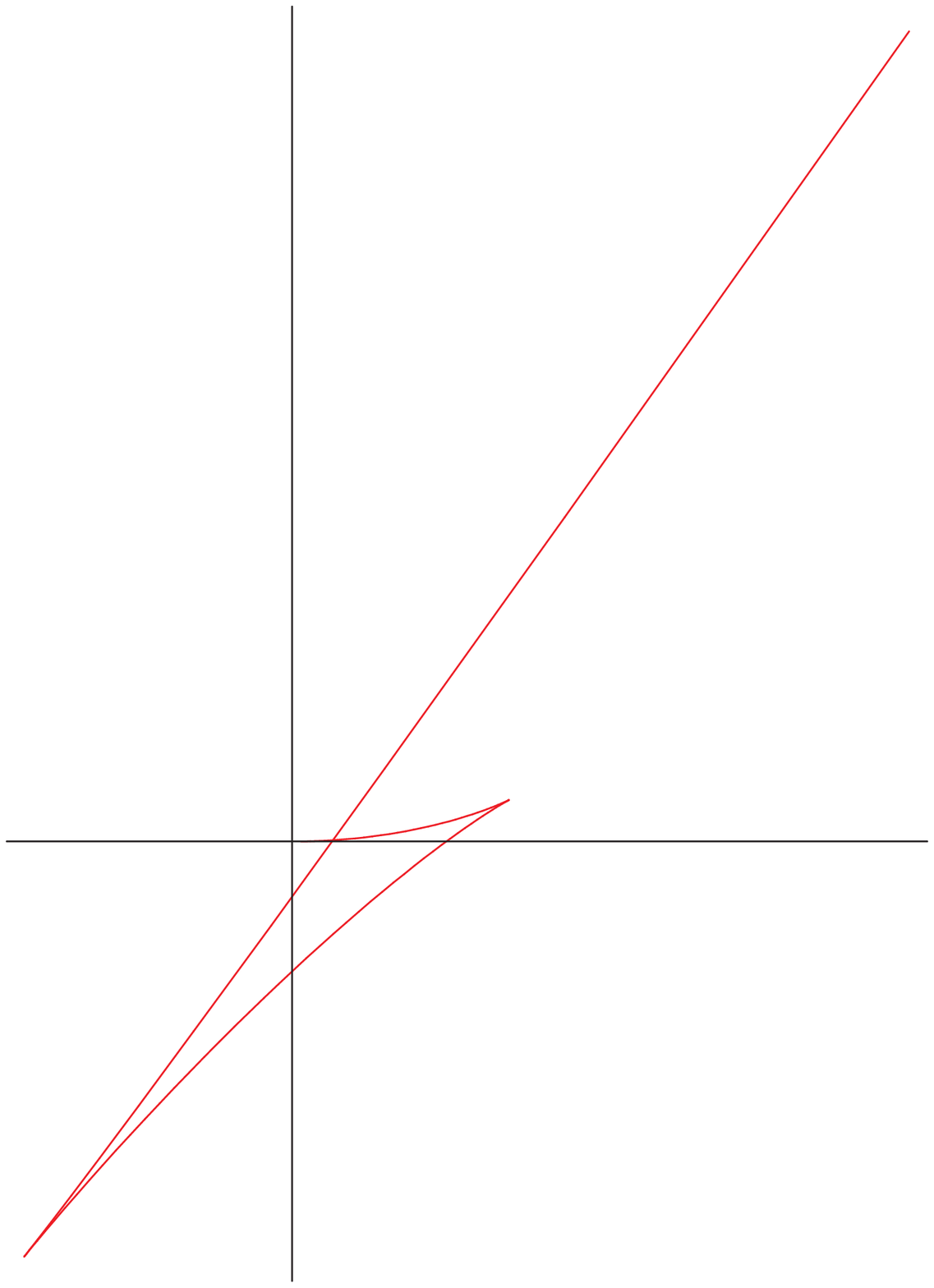,height=2in}
         \hskip4.0cm
         \psfig{figure=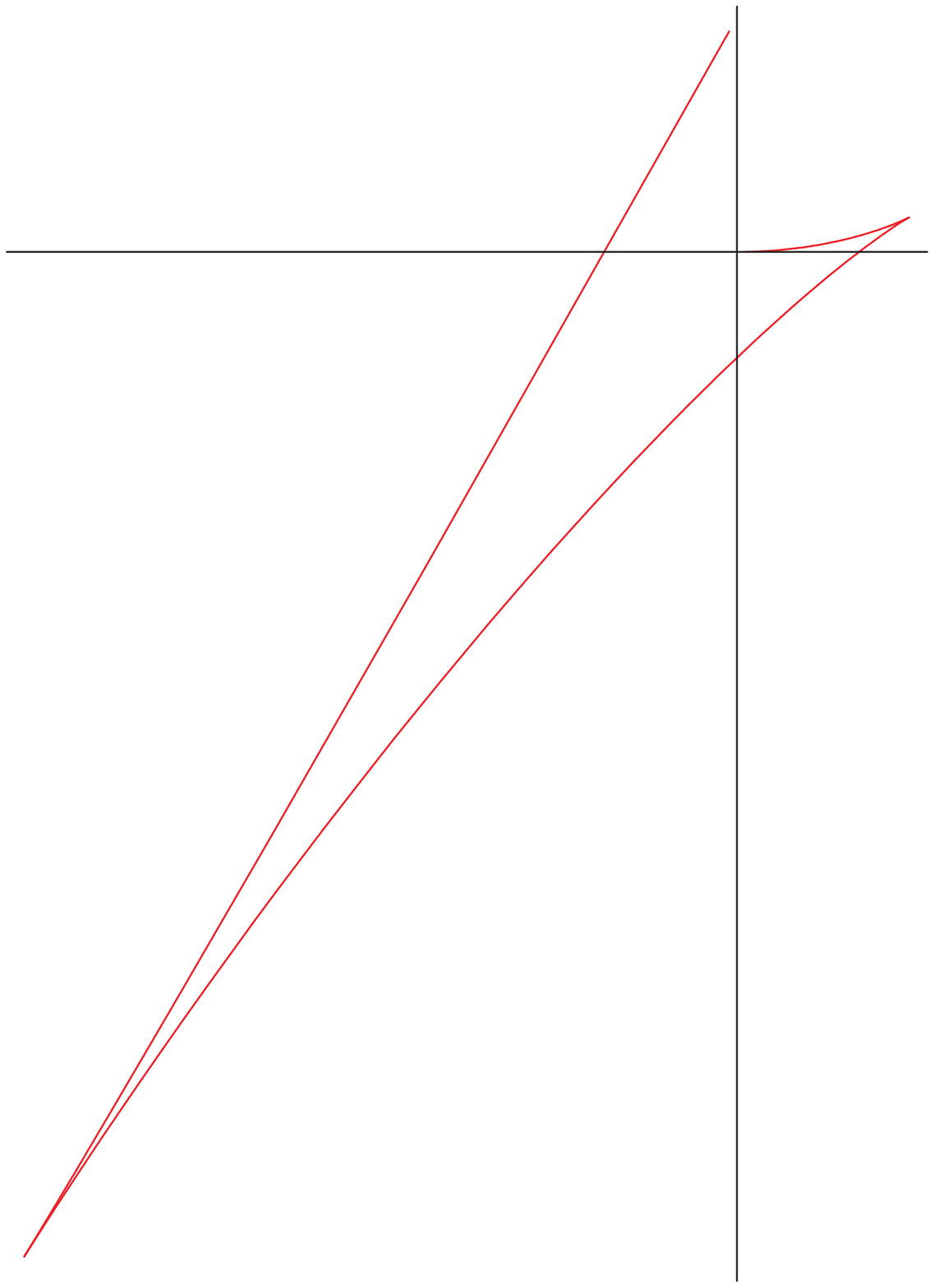,height=2in}  }
         \vskip0.5cm
         {\footnotesize Figure 8.  The top left-hand plot shows
         $L(t)$ for the near extremal black hole, where there is no
         transition.  The other plots show the case of first order
         behaviour discussed in the text, which exists far enough from
         extremality.  The black hole is further from extremality as
         the plots are read from top left to top right, to bottom left
         to bottom right.}
\end{figure}

Suppose at the energy $E_{\rm crit}$ the same branch of geodesics
has the least action.  In that case there will be no source of
non-analyticity at $t_{\rm crit}$, as the length of the geodesic
(and therefore the value of the correlator) will be a smooth
function of the boundary point.  It may then be that at some lower
value of $t$ one of the other branches will have the least action.
This will generally result in non-analyticity similar to a
second (or higher) order phase transition -- the value of the
correlator will be continuous past the transition, but some (time)
derivative thereof will be discontinuous.

On the other hand, behaviour analogous to a first order transition
will result if one of the branches which develops at $t_{\rm crit}$ immediately
dominates the correlation function.  There is then no reason for the
lengths of the geodesics on both branches to be degenerate, and in general
one will discover a discontinuity of the correlation function, as
opposed to its derivatives.

For the values of the parameters we plot in figure 8, we see that
when the transition exists it is always of first order, yielding a
discontinuity in the correlator.  By continuity it is clear that by
tuning   parameters we can make the two branches  degenerate; thus
as we vary the black hole parameters  we obtain a line of first
order transition terminating at a point of  second order transition.


\sect{Charged Correlation Functions}

We now discuss correlation functions of charged operators. As the
analytic expressions are significantly more involved, we confine
ourselves to discussing the qualitative features of the correlators,
which are fairly similar to the neutral case discussed above.

To approximate correlators of operators which carry R-charge, we
need to discuss trajectories of electrically charged
particles\footnote{Such particles exhibit interesting phenomena such
as Schwinger pair production and induced emission~\cite{gibbons}.
See, e.g.,~\cite{affleck} and, for a recent
discussion,~\cite{us}.  The Schwinger effect in curved space is
studied in~\cite{herman}.}.  We will only consider radial motion here, the
addition of angular momentum having much the same effect as in the
uncharged case\footnote{As in that case, for an arbitrarily small
  angular momentum, one can find trajectories probing the region beyond
  the Cauchy horizon, but they return to a copy of the boundary on the
  same side of the Penrose diagram. These are therefore irrelevant to
  the two-sided correlators discussed here, and more generally we
  expect such trajectories not to dominate any gauge theory correlator.}. The
Lagrangian describing such a particle with charge $q$ and unit mass
is
\be
{\cal L} = \frac{1}{2} g_{\mu \nu} \dot{X}^\mu
\dot{X}^\nu + q A_\mu \dot{X}^\mu = \frac{1}{2}
\left( -f(r) \dot{t}^2 +
  \frac{\dot{r}^2}{f(r)} \right)  - \frac{\Q}{r^2} \dot{t}
\ee where we have defined $\Q = cqQ$.

The worldline Hamiltonian $\mathcal{H} = (-f(r) \dot{t}^2 +f^{-1}(r)
\dot{r}^2)/2$ is conserved, since the action is time translation
invariant on the worldline. As we are interested in spacelike paths
we can rescale $\lambda$ to set $\mathcal{H}=1/2$. The equations
of motion are then
\be \label{eqn}
\dot{t} = \frac{1}{f(r)} \left( E - \frac{\Q}{r^2}
\right), \qquad \dot{r}^2 = \left( E -
\frac{\Q}{r^2} \right)^2 + f(r).
\ee
Note that shifting $A_t$ by a constant (denoted in section 1 by $\Phi$), which
is just a gauge transformation, uniformly shifts the energy of all
trajectories. This gauge freedom was fixed above, so that $A_t$ vanishes at the
outer horizon. We choose here to absorb the constant part of $A_t$, into
the definition of $E$.

The relations following from (\ref{eqn}) are as follows. First, the
(real part of) the boundary time, for symmetric trajectories, is
\be
\label{time}
t_0  =
\int_{r_E}^\infty \d r \frac{(E- \frac{\Q}{r^2})}{V(r)
\sqrt{(E - \frac{\Q}{r^2})^2
-V(r)}}
\ee
where $r_E$ is the turning point described below. The
length of the corresponding trajectory is given by
\be
\label{length1}
L = 2 \int_{r_E}^{r_{\rm max}} \frac{\d r}{\sqrt{(E- \frac{\Q}{r^2})^2 -
V(r)}}
\ee
where $r_{\rm max}$ is the long distance bulk cutoff introduced in
section 3.

The equation for $r(\lambda)$ given in (\ref{eqn}) cannot be
interpreted as
describing a particle with energy $E^2$ moving in a potential $V(r)$,
since now the potential itself depends on $E$. However, one can still
describe the qualitative features of the trajectory,
which now depend on both the parameters $\Q$ and $E$.

The turning point, $r_E$, of the trajectory is the largest root of
the equation \be \left( E - \frac{\Q}{r^2} \right)^2 - V(r) = 0. \ee
To visualise the situation, in figure 9 we plot $V(r)=-f(r)$ and the
(positive semi--definite) function $(E - \Q / r^2)^2$, the behaviour
of which depends on the sign of $\Q/E$.  We therefore plot the
function for both signs.

\vskip 0.5cm\begin{figure}[ht]
         \beginlabels\refpos 0 0 {}
                     \put 180 -160 {r}
                     \put 428 -160 {r}
                     \put 30 -150 {E^2}
         \endlabels
         \centerline{
         \psfig{figure=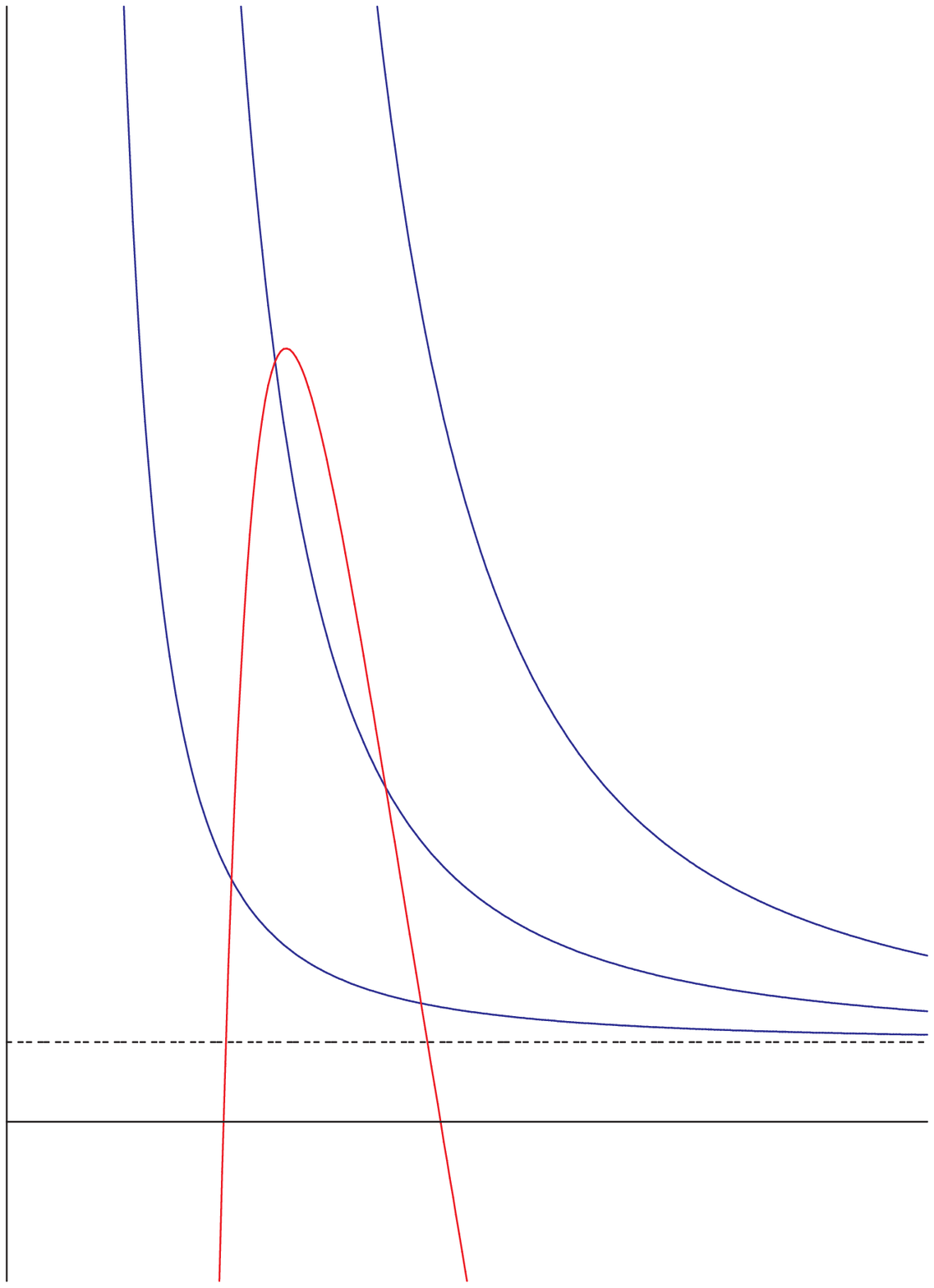,height=2.5in}
         \hskip4.0cm
         \psfig{figure=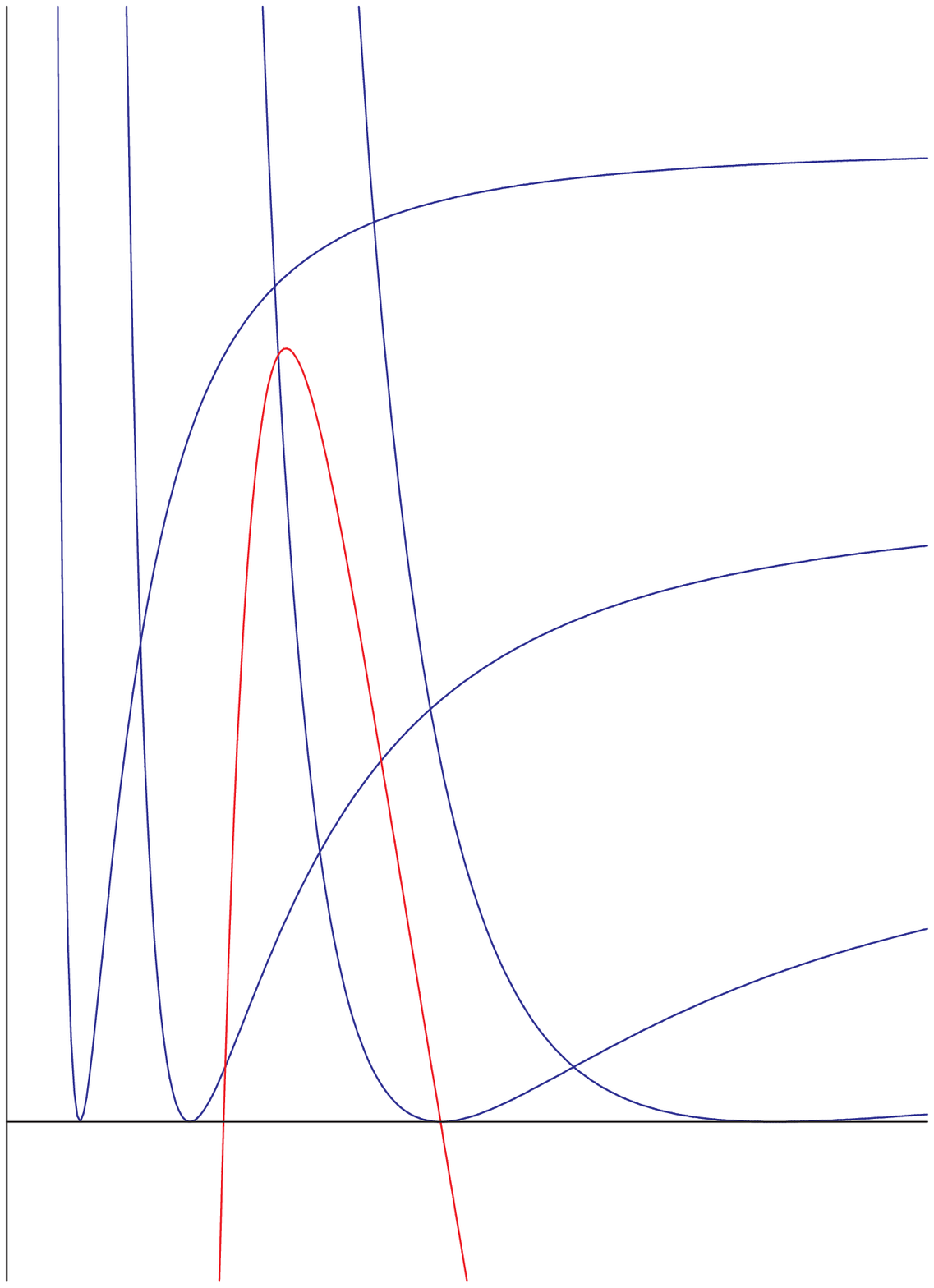,height=2.5in} }
         \vskip0.5cm
         {\footnotesize Figure 9.  The left-hand plot has $\Q/E<0$.
         $V(r) = -f(r)$ is in red
         and $(E - \Q / r^2)^2$ in blue.  Three trajectories with the
         same energy, $E^2$, but different
         charges, are shown.  The right-hand plot has $\Q/E>0$, and shows four
         trajectories with various values of the parameters.  One has
         $\Q/E > r_+^2$ without a radial turning point, one has $\Q/E
         = r_+^2$ and two have $\Q/E < r_-^2$, one with a radial
         turning point and one without.}
\end{figure}

In either case, there are three possibilities:
\begin{itemize}

\item The graphs do not intersect, and there is no classical turning
  point. These classical trajectories end at the singularity.

\item The graphs intersect twice, in which case the turning point is
  that intersection with the largest value of $r$.  As in the
  uncharged case, this will always be in the region between the inner
  and outer horizons.

\item The borderline case, in which there is a single intersection
  point, now not necessarily at precisely the top of the potential.
  Similar to the  discussion above,
  looking at the behaviour of (\ref{time}) and (\ref
  {length1})
  near $r_E$  reveals   logarithmic divergences both in the boundary
  time and in the proper proper length of the geodesics. Therefore
  this value of $E$ should be viewed as a limiting value, trajectories
  with larger value of $E$ being irrelevant for our purposes.

\end{itemize}

These features of $r(\la)$ are similar to those discussed
previously.  It is clear that only those trajectories with a radial turning
point will return to a boundary, but it is not clear which
boundary this will be.  To determine which trajectories are relevant
for our two-sided correlators, we must also consider the behaviour of
$t(\la)$, which is more involved than in the uncharged case.  Let us
take $E > 0$ so that $t$
increases with affine parameter from the initial
boundary\footnote{Trajectories with $E<0$, but the same sign of
$\Q/E$,  will just be mirror
images of the trajectories we consider.}.  Then, for $\Q < 0$
(the particle having a charge of opposite sign to
the black hole), $\dot{t}$ just depends on the sign
of the function $f(r)$: for $r>r_+$ it is positive,
whereas for $r_- < r < r_+$ it is negative.  So, if
such trajectories have a radial turning point, and
they do for small enough $\Q$, then they
\emph{will} connect the two boundaries of interest. In this case
the coordinate $t(\la)$ is similar to the neutral geodesics.

The situation for $\Q > 0$ (the
particle having a charge of the same sign as the
black hole), is more complicated. The behaviour depends
on the magnitude of
$\Q/E$, and in each case is given by inspection
of the right-hand side of the equation for $t(\la)$ in (\ref{eqn}):
\begin{itemize}

\item For $\Q/E > r_+^2$, there is some $r>r_+$ for which $\dot{t}=0$,
and the trajectory will start to move backward in
coordinate time before reaching the outer horizon.
It crosses the event horizon, upon which it
moves forward in coordinate time again.  If such
trajectories have a radial turning point (and some
certainly do), then they will connect the two
boundaries of interest.

\item For $\Q/E = r_+^2$, $\dot{t}=0$ precisely at the outer horizon,
where the particle will necessarily have a radial
turning point.  It will then return to the same
boundary, so such trajectories are of no concern to
us.

\item For $r_-^2 \le \Q/E < r_+^2$, $\dot{t}$ does not vanish unless
  $r_- < r < r_+$.  However, it
necessarily has a radial turning point
\emph{before} the radius at which $\dot{t}=0$, so
will indeed connect the two boundaries of interest.

\item Finally, for $\Q/E < r_-^2$, the particle may or may not have a
radial turning point.  If it does, then it will
connect the boundaries of interest.  Otherwise, the
sign of $\dot{t}$ will flip at some $r <r_-$, and the trajectory will
always hit the singularity.

\end{itemize}

The qualitative behaviour of the charged trajectories which do connect the
two boundaries is thus
similar to the neutral spacelike geodesics analysed above.  They never
cross the Cauchy horizon and, again, only reach a radius strictly
larger than $r_-$.


\sect{Scanning Behind the Horizon}

\subsection{Perturbing the AdS-RN Spacetime}

When the charged black hole is perturbed, linearised analysis
suggests strong back-reaction near the inner
horizon~\cite{penrose:68,matzner,chan:82}. This is one instance of
physics behind the horizon which would be nice to interpret in the
gauge theory.  We take here some preliminary steps towards this
goal, postponing a more complete discussion for~\cite{progress}.

First, if one restricts attention to null perturbations which are
either purely ingoing, or purely outgoing\footnote{ For a massless
field in AdS space, there are two types of perturbations, the
non-normalisable one goes to a constant near the boundary. These are
the type of perturbations we consider here. In the gauge theory this
corresponds to turning on a marginal perturbation, with a specific
time-dependent profile.} (i.e., they are chiral in the sense of
\cite{chicago}), then one can solve for the perturbed metric
exactly~\cite{poisson,progress}. One obtains an AdS version of the
charged Vaidya solution~\cite{vaidya,poisson}. This solution
contains one arbitrary function (the mass function) which depends on
the profile of the perturbation. That profile can be varied, and can
be arbitrarily localised\footnote{This
  arbitrariness is similar to gravitational plane waves which solve
  Einstein's equation for an arbitrary wave profile.}. By investigating the
two-sided correlators (and other observables) as a function of the
perturbation, one obtains an efficient method of scanning (at least
some of) the region behind the horizon. We will demonstrate here the effects on
our observables of a simple wave profile, that of an infinitely
localised delta function pulse.

Secondly, in the asymptotically flat case, when both types of null
perturbations are turned on\footnote{These
are still not generic perturbations, so cosmic censorship considerations
do not apply.} the spacetime is drastically changed near the inner
horizon, resulting in the phenomenon of {\it mass
  inflation}~\cite{poisson}.  We expect that this phenomenon persists
in the asymptotically AdS case~\cite{mann,progress}.  We
further expect that the dependence of some gauge theory observables
will be non-analytic
as a function of the perturbation strength: taking that strength
to be arbitrarily weak does not diminish its effect.

Such non-analytic behaviour is not uncommon in field theories, and
results from the existence of infra-red divergences. When re-summing
perturbation theory, one discovers non-analytic dependence on coupling
constants. Such behaviour would be slightly surprising for gauge theory
on a compact space (a three-sphere in this case), at finite
temperature and density, but it may be sensible in the infinite N limit.

Here, we find a behaviour which is perhaps more sensible. As we saw
above the spacelike geodesics accumulate at some finite distance
from the Cauchy horizon. This is no longer necessarily true of the
perturbed geometry. Nevertheless the geodesics we investigate seem
to always be screened from the dramatic behaviour at the Cauchy
horizon, as is demonstrated below. It is unclear   if this is a
property of the particular observables we are studying, or a more
general property of the gauge theory.


\subsubsection*{An Asymptotically AdS Charged Vaidya Solution}

The Vaidya solution~\cite{vaidya} describes an ingoing or outgoing
null shell of matter incident on a vacuum black hole.  Poisson and
Israel generalised this to include charge in~\cite{poisson}, their
aim being to investigate the properties of the Cauchy horizon of the
asymptotically flat Reissner-Nordstr\"{o}m black hole. It is easy to
further generalise these solutions to the asymptotically AdS case.
In fact, though we will not do this here, one can
derive~\cite{progress} asymptotically AdS versions of the solutions
relevant to mass inflation~\cite{poisson}, describing both ingoing
and outgoing flux.

In terms of the lightcone coordinate $u=t-r_*$ defined in section 2, the
outgoing solution, in the region $r_- < r < r_+$, has the metric
\be
\d s^2 = - f(u,r) \d u^2 - 2 \d u \d r + r^2 \d \Om^2_3
\label{eqn:metric}
\ee
where all the $u$-dependence in the metric function is through that of the
mass function $M(u)$: \be\label{func} f(u,r) = 1 - \frac{M(u)}{r^2}
+ \frac{Q^2}{r^4} + r^2. \ee The $u$-dependence is set by the flux
sent in from the left-hand boundary, which must have the null
energy-momentum tensor, whose only non-zero component is\be T_{uu} =
\frac{1}{2} \frac{\del_u M(u)}{r^3}. \ee

 It is easy to check that this
metric and energy-momentum tensor, together with the background
gauge field given in (\ref{eqn:soln}) solve the five dimensional
Einstein-Maxwell field equations with negative cosmological
constant.


\subsubsection*{A Simple Example}

The simplest case, which we will consider here, is to take
\be
M(u) = m_1 + \Theta(u-u_p) \Delta m.
\label{eqn:mass}
\ee
The corresponding energy-momentum tensor
\be
T_{uu} = \frac{\Delta m}{2} \frac{\delta(u-u_p)}{r^3}
\ee
corresponds to an infinitely thin shell
of null matter, the flux leaving the left-hand
boundary at $t=t_p$.  The spacetime splits into two regions: for
$u<u_p$ the mass is $m_1$ and for $u>u_p$, the mass is $m_2 = m_1 +
\Delta m$.  We can then apply the previous analysis to the two
regions, each with its own potential, and match across the $u=u_p$
surface. For non-rotating neutral geodesics the potential in each
region is simply $V(r)= - f_{m_{1,2}}(r)$, where $f_m(r)$ is the
radial function (\ref{func}) with specific value of the mass
parameter $m$.

If we take $\Delta m > 0$, then $f_{m_2} < f_{m_1}$ for all values
of $r$. The roots $x_-(m_2)$ and $x_+(m_2)$ of $f_{m_2}$ will thus
be respectively less than $x_-(m_1)$ and greater than $x_+(m_1)$.
The resulting geometry is shown schematically in figure 10.

\vskip 0.5cm\begin{figure}[ht]
         \beginlabels\refpos 0 0 {}
                     \put 250 0 {{\rm Inner~apparent~horizon}}
                     \put 115 -62 {{\rm Cauchy}}
                     \put 115 -72 {{\rm horizon}}
                     \put 341 -83 {r_I,\,t_I}
                     \put 118 -237 {t_p}
                     \put 180 -300 {{\rm Event~horizon}}
         \endlabels
         \centerline{
         \psfig{figure=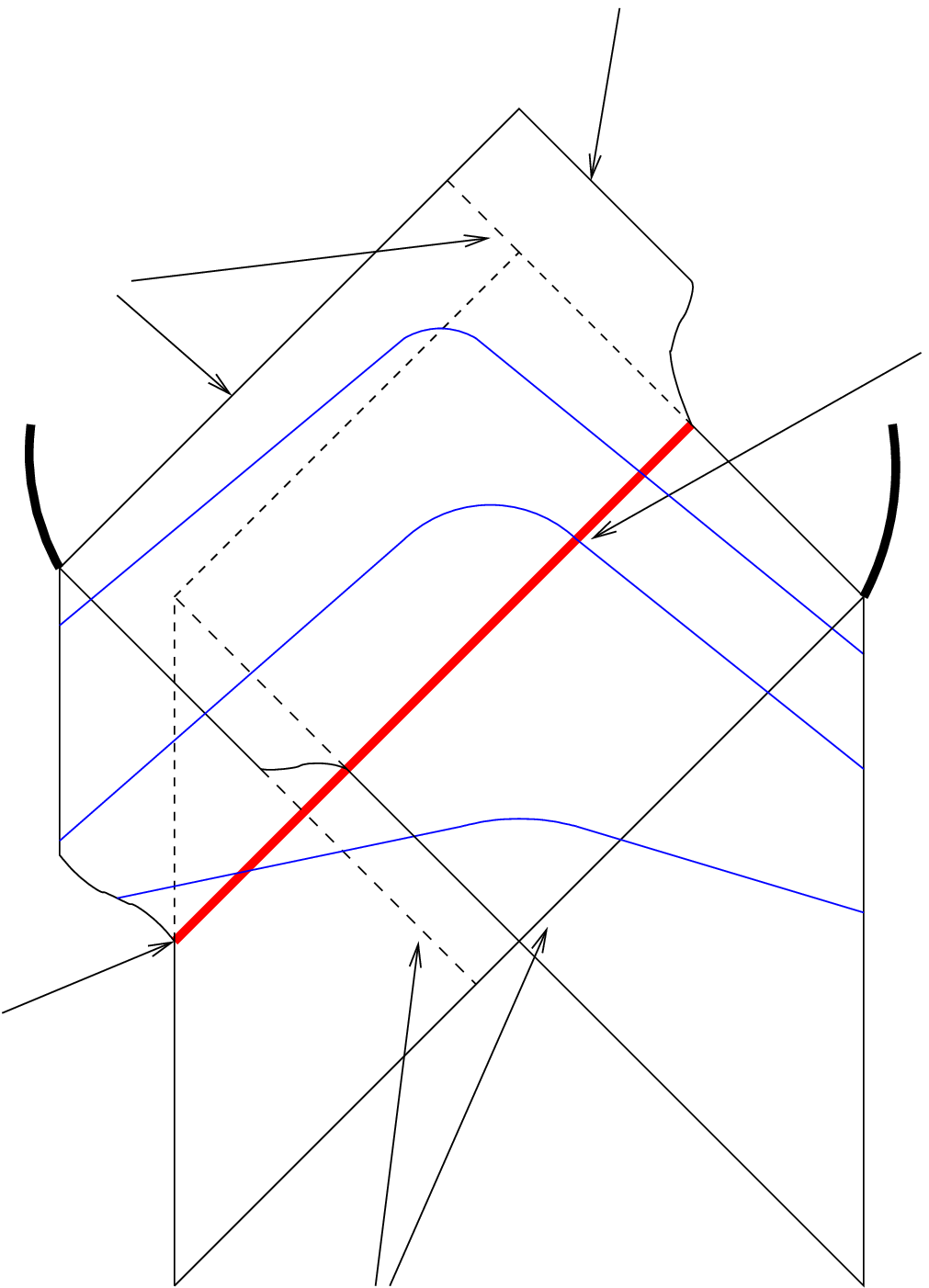,height=4.0in} }
         \vskip0.5cm
         {\footnotesize Figure 10.  The geometry of the perturbed
         spacetime, showing the event horizon, the Cauchy horizon and
         the inner apparent horizon.  The perturbation leaves the
         left-hand boundary at $t=t_p$, and is shown as a solid red
         line.  Spacelike geodesics are shown schematically in blue,
         the point at which they interact with the perturbation being
         shown as $r_I, t_I$.}
\end{figure}

Now, to see the effect of the perturbation on the gauge theory,
consider again the two-sided correlator in the geodesic
approximation. The perturbation is then encoded in properties of
spacelike geodesics which intersect the surface $u=u_p$. Some
possible geodesics are shown schematically in figure 10, in which we denote
the intersection point by $r_I, t_I$.  The interaction with the flux
will change the path of the geodesic beyond this intersection point.
We are only interested in those geodesics which intersect the flux
in the region beyond the outer horizon and it is clear that
late-time correlators will be most suitable.

As we change the time $t_p$ at which the pulse leaves the boundary,
the intersection point will change, and so will the length $L$ of
any  geodesic. We should thus be able to see how the correlator
changes as a function of $t_p$, a clear signal of physics behind the
horizon.  We demonstrate here how the information
about the perturbation can be detected in our simple example.

The only subtlety\footnote{As pointed out to us by Simon Ross.} is
that the energy of the geodesics is no longer conserved as they cross
the null shell.  In the first region, from the metric
(\ref{eqn:metric}), the conserved quantity is
\be
E = -\frac{1}{2} \frac{\del {\cal L}}{\del \dot{u}} = f(r) \dot{u} +
\dot{r}
\ee
which is equal to the conserved energy $E=-\frac{1}{2} \frac{\del
  {\cal L}}{\del \dot{t}}$ in the $t,r$ coordinates we have been using for the
unperturbed geometry.  In the perturbed case, however, $E$ is no longer
conserved: if we define $E\equiv f(r,u) \dot{u} + \dot{r}$
then, from the geodesic equations,
\be
\frac{\d E}{\d \la} = \frac{1}{2} \del_u f(r,u) \dot{u}^2
\ee
which, for the mass function as in (\ref{eqn:mass}), gives
\be
\d E = -\frac{1}{2} \frac{\Delta m}{r^2} \delta (u-u_p)\,
\dot{u} \,\d u.
\ee
Whereas the ``energy'' $E$ is discontinuous across the null shell, the
momentum normal to the shell, which is just $p_\perp = \dot{u}$, \emph{is}
continuous at the interaction point~\cite{barrabes}.  We can thus
integrate the above equation, taking $\dot{u} = \dot{u}(m_1)$, the
value of the momentum in the region with mass $m_1$.  From the
definition of $E$, and
using the Lagrangian constraint ${\cal L} = 1$ for spacelike geodesics, we have
\be
\dot{u} = \frac{E}{f(r,u)} \left( 1 \pm \sqrt{1 + \frac{f(r,u)}{E^2}}
\right),
\ee
so that
\be
E_2 - E_1 = -\frac{1}{2} \frac{\Delta m}{r_I^2} \frac{E_1}{f_{m_1}(r_I)}
\left( 1 \pm \sqrt{1 + \frac{f_{m_1}(r_I)}{E_1^2}} \right)
\label{eqn:energy}
\ee
where $E_1$ and $E_2$ denote the respective energies in the regions
with mass $m_1$ and $m_2$.  Note that
$f(r_I,u_p) < 0$ so that $E_2 > E_1$.


\subsection{Detecting the Perturbation}

\subsubsection*{Qualitative Features}

To analyse the qualitative properties of the geodesics, we
re-consider the particle mechanics problem of section 3.  The
particle moves from the right-hand asymptotic region to the
intersection point at $r=r_I$ in the effective potential $V_{m_1}(r)
= -f_{m_1}(r)$, and with energy $E_1$.  After the intersection point,
the particle
continues on its motion, but now in the effective potential
$V_{m_2}(r) = -f_{m_2}(r)$, and with energy $E_2$. The various
possibilities are shown in
figure 11, in which we fix the energy of the geodesic, and vary the
mass, which in turn changes the potential.  We can vary the
intersection point by varying $t_p$. There are then three
possibilities (we will assume that the particle continues its motion
entirely in $V_{m_2}$ after the intersection, i.e. that it does not
intersect the pulse a second time):
\begin{itemize}

\item If the intersection point satisfies $r_I>r_E(m_2)$, the turning point of the
motion in $V_{m_2}(r)$, then the particle continues its motion in
the modified potential until reaching the new  turning point
$r_E(m_2)$, and will thus still return to the opposite boundary.
This is shown in the left-hand plot in figure 11.

\item The centre plot has $r_I<r_E(m_2)$, so that after the
         interaction, it seems that the particle would be moving \emph{below}
         $V_{m_2}(r)$.  However, the new energy in the second region
         is $E_2 > E_1$, and this
         shift in the energy across the null pulse is sufficient to
         cause the particle to jump back above the second
         potential (we have checked that the shift in energy is large
         enough for a range of
         black hole parameters).  This case is thus
         qualitatively similar to the previous one.

\item The right-hand
         plot shows two ingoing particles, both with $E^2_1 > V_{\rm
         max}(m_1)$, so they would have fallen into the singularity
         in the unperturbed case.  Here, however, they still have $r_I
         > r_E(m_2)$, so will continue their motion until reaching
         this turning point.  Contrary to the unperturbed case, such
         geodesics will thus return to the opposite boundary.

\end{itemize}

The last case is interesting, since the trajectories now seem to cross the
Cauchy horizon.  However, as shown schematically in figure 10, by
studying the trajectories more carefully one can see
that this does not really happen.  The trajectories cross the
left-hand branch of the surface
$r_-(m_1)$, which would have been the Cauchy horizon in the
unperturbed spacetime, but is not in the perturbed case.  The
geodesics never cross the surface $r_-(m_2)$, which is the left-hand
branch of the Cauchy surface in this case.  Once again, it
seems that the correlators are geometrically protected from the
catastrophic instability of the Cauchy horizon.

\vskip 0.5cm\begin{figure}[ht]
         \beginlabels\refpos 0 0 {}
                     \put 85 -140 {r_I}
                     \put 250 -163 {r_I}
                     \put 378 -185 {r_-}
                     \put 412 -185 {r_I}
                     \put 0 0 {V}
                     \put 162 0 {V}
                     \put 326 0 {V}
                     \put 140 -67 {E^2_1}
                     \put 300 -118 {E^2_1}
                     \put 463 -37 {E^2_1}
                     \put 463 -118 {E^2_1}
                     \put 140 -130 {r}
                     \put 300 -152 {r}
                     \put 463 -173 {r}
         \endlabels
         \centerline{
         \psfig{figure=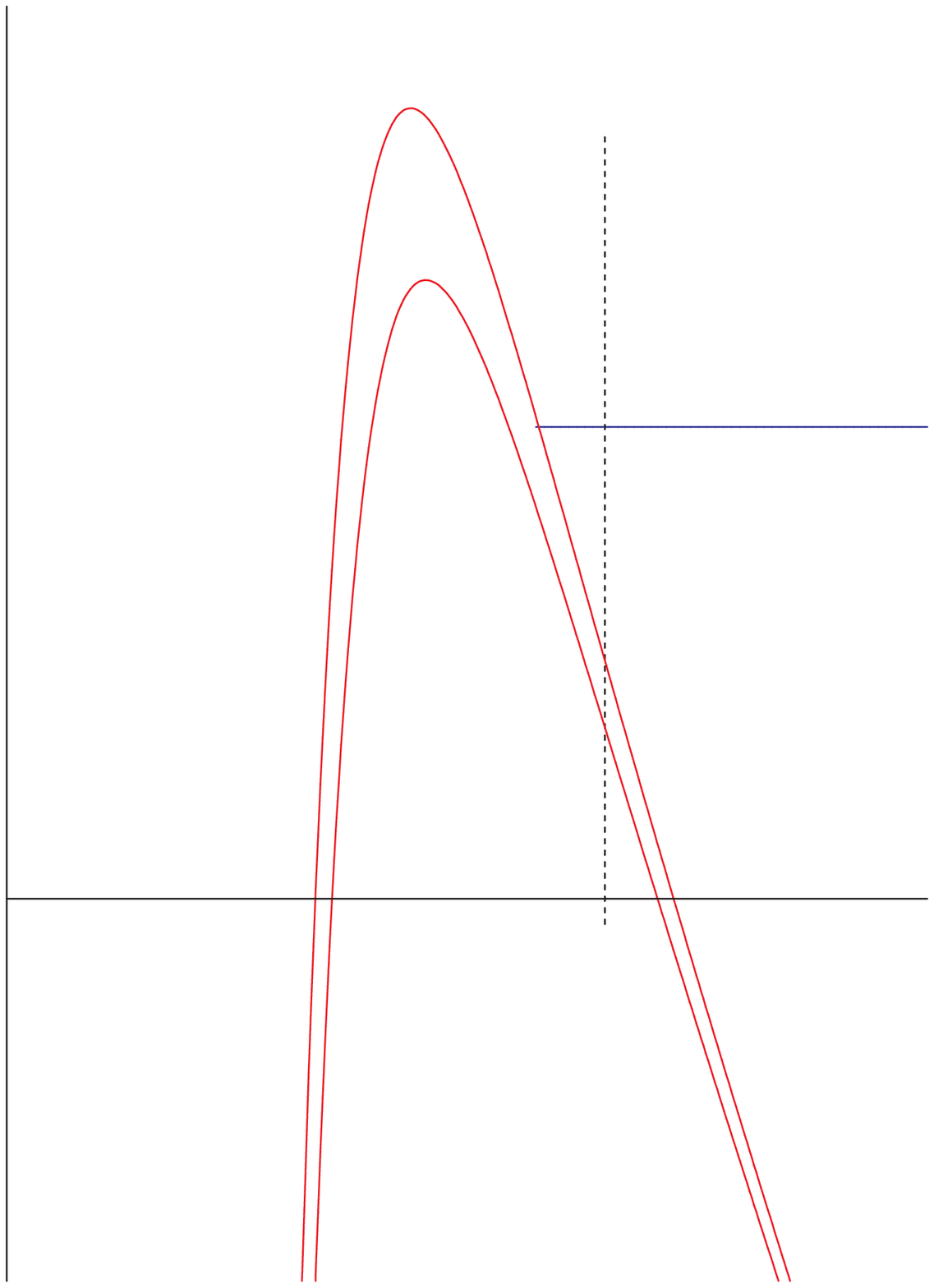,height=2.5in}
         \hskip1.0cm
         \psfig{figure=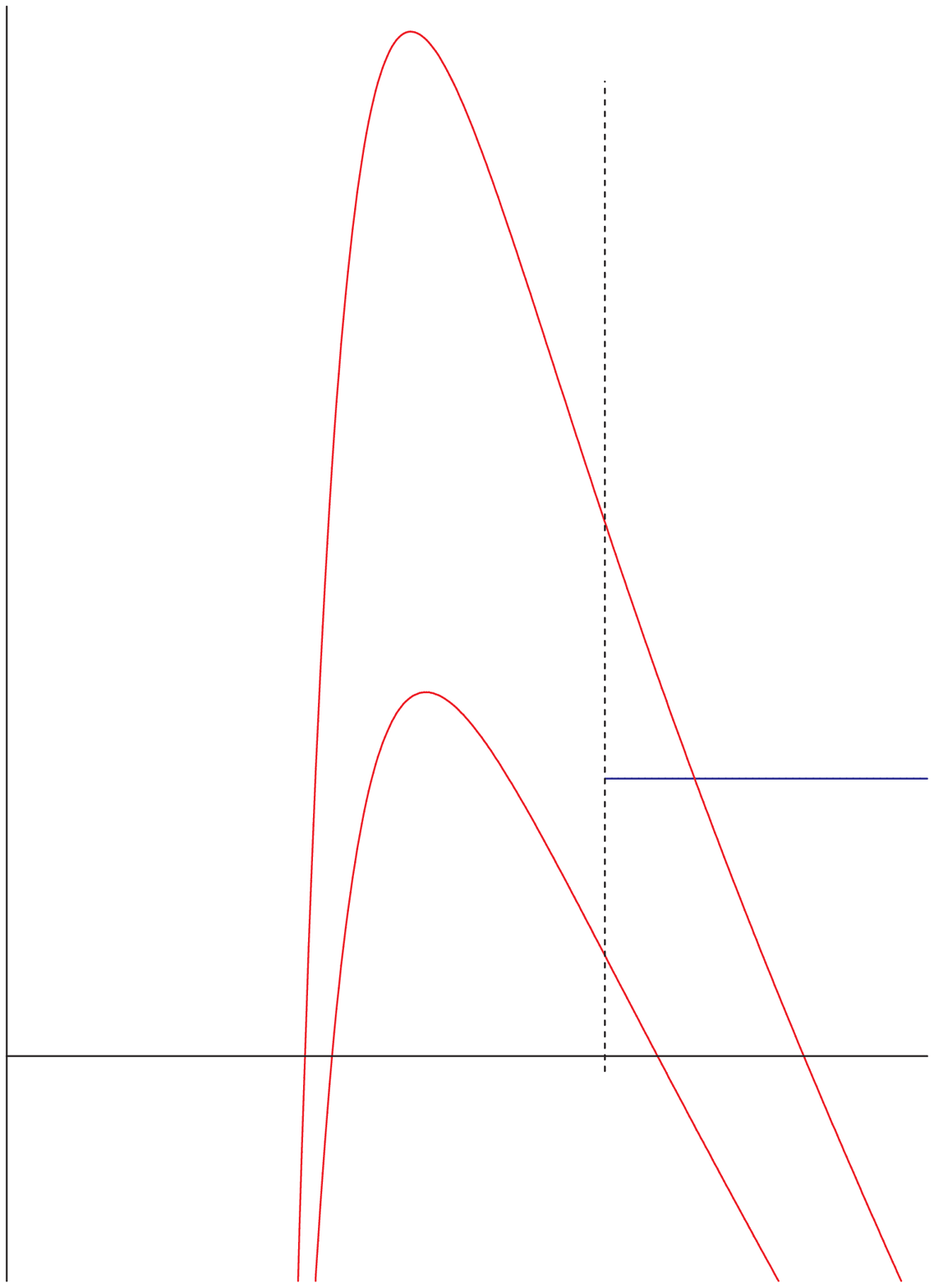,height=2.5in}
         \hskip1.0cm
         \psfig{figure=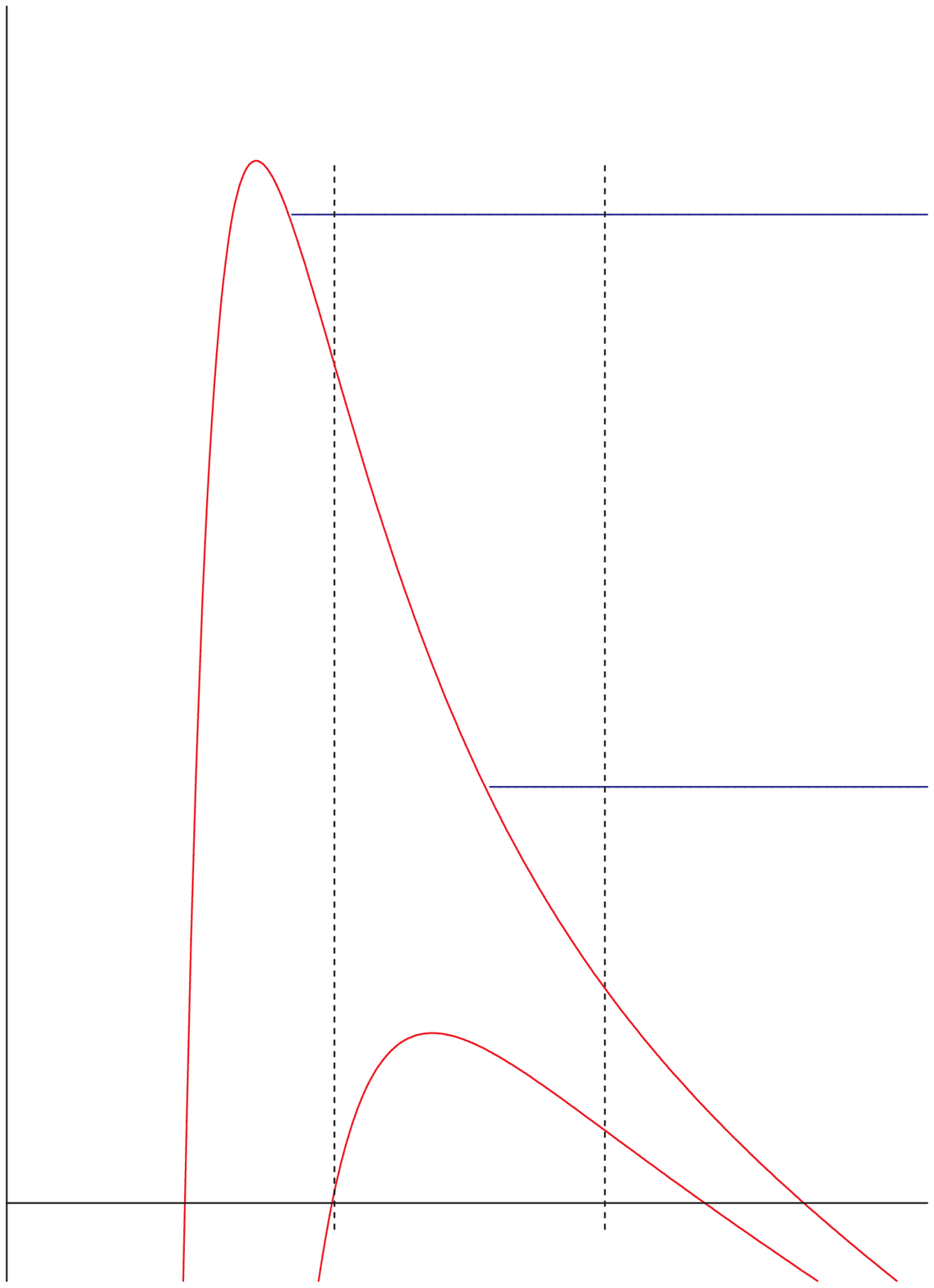,height=2.5in} }
         \vskip0.5cm
         {\footnotesize Figure 11.  The
         particles with energy $E^2_1$ are initially moving in the
         smaller potential,
         $V_{m_1}(r)$.  They interact with the perturbation at some $r=r_I
         < r_+(m_1)$, inside the outer horizon, at which point the
         energy jumps to $E^2_2 > E^2_1$, and they
         continue to move in the
         larger potential, $V_{m_2}(r)$.  The various possibilities shown
         are discussed in the text.}
\end{figure}

To demonstrate  the scanning process of the region behind the event
horizon, we will study the features of geodesics which fall into the
first class discussed above, those with $E^2_1 < V_{\rm max}(m_1)$
which intersect the flux at $r_I > r_E(m_2)$. Such geodesics connect
both boundaries of the perturbed spacetime, and therefore dominate
the correlation functions of the perturbed gauge theory.


\subsubsection*{A Quantitative Plot of the Correlators}

It is difficult to solve analytically for the behaviour
of the correlator in the perturbed spacetime. Instead we demonstrate
the process by numerically plotting the length of the geodesic as a
function of $t_p$, the time of the perturbation.

Specifically , we  plot here the fractional change in the length of
the geodesic as a function of $t_p$,
\be
\delta L = \frac{L(\Delta m) - L(\Delta m = 0)}{L(\Delta m =0)}.
\label{eqn:deltaL}
\ee
The method to generate the plot is not entirely obvious, so we outline it
before presenting the result.

We first fix the initial roots, $x_\pm(m_1)$, of the potential
$V_{m_1}(r)$, which fixes the initial mass $m_1$ and the charge
(which remains unchanged).  We fix $\Delta m$, the strength of the
perturbation, which fixes the final mass $m_2$, and also the new
roots $x_\pm(m_2)$ of the potential $V_{m_2}(r)$. Finally, we fix
the initial energy $E^2_1$ of the geodesic in the unperturbed geometry, which
fixes the boundary time of
the correlator in that case. We take $E^2_1 \sim V_{\rm max}(m_1)$ to give late
time geodesics, which should be guaranteed to interact with the
perturbation in the relevant region.

In the unperturbed spacetime, with mass $m_1$, the geodesic with
energy $E^2_1$ connects two specific points on opposite boundaries.
The proper length of this approximates the correlator of operators
inserted at these points.  We need to compare this with the correlator
in the perturbed spacetime, which \emph{connects the same
  two points}.  The correlators we compute before and after perturbing
the spacetime should be the same; in the geodesic approximation,
however, they are dominated by geodesics with different energies,
say energy $\tilde{E}^2_1$ in the first region of the perturbed geometry (and,
as discussed above, with a shifted energy $\tilde{E}^2_2$ in the
second region).

Let us denote the points on the left- and right-hand boundaries
which the geodesics connect as $t_l$ and $t_r$.  In the unperturbed
spacetime, we have
\be
t_l(m_1) + t_r(m_1) = 2 E_1
\int_{r_{E_1}(m_1)}^\infty \frac{\d r}{V_{m_1}(r)
  \sqrt{E^2_1 - V_{m_1}(r)}},
\label{eqn:t1}
\ee
which can be computed as in subsection 3.2.  This
must be equal to the same quantity in the perturbed spacetime:
\[
t_l(m_2) + t_r(m_1) = \tilde{E}_1 \int_{r_{\tilde{E}_1}(m_1)}^\infty
\frac{\d r}{V_{m_1}(r)
  \sqrt{\tilde{E}^2_1 - V_{m_1}(r)}} - \tilde{E}_1 \int_{r_{\tilde{E}_1}(m_1)}^{r_I} \frac{\d
  r}{V_{m_1}(r) \sqrt{\tilde{E}^2_1 - V_{m_1}(r)}}
\]
\be + \tilde{E}_2
  \int_{r_{\tilde{E}_2}(m_2)}^\infty \frac{\d r}{V_{m_2}(r)
  \sqrt{\tilde{E}^2_2 - V_{m_2}(r)}} + \tilde{E}_2 \int_{r_{\tilde{E}_2}(m_2)}^{r_I} \frac{\d
  r}{V_{m_2}(r) \sqrt{\tilde{E}^2_2 - V_{m_2}(r)}}
\label{eqn:t2}
\ee
as can be seen by inspection of figure 10, and where $\tilde{E}_2$ is
related to $\tilde{E}_1$ as in (\ref{eqn:energy}). Again,
we can compute this in terms of elliptic integrals as in subsection
3.2.  In the perturbed case the result is a function of the
intersection radius $r_I$.  Equating the two expressions for the
boundary time (\ref{eqn:t1},\ref{eqn:t2}) in principle gives then a relation
$r_I(\tilde{E}_1)$.  In practice, however, we need to know the
interaction radius to compute $\tilde{E}_2$ (through
(\ref{eqn:energy})), which in turn we need to compute the interaction radius.

We can avoid this vicious circle by running through values of $r_I$
(knowing that $r_+(m_1) > r_I > r_-(m_2)$)
for each $\tilde{E}_1$, computing $\tilde{E}_2$ at each step (we take
the smaller solution, with the negative sign in (\ref{eqn:energy}), to
ensure $\tilde{E}_2^2 < V_{\rm max}(m_1)$),
then comparing the two expressions (\ref{eqn:t1},\ref{eqn:t2}) for
these specific values.  If they are close enough, then we know that we
have the correct value for $r_I$.  Numerical experiment for various
values of the black hole parameters shows that the
difference between the two expressions (\ref{eqn:t1},\ref{eqn:t2}) for
the boundary time is initially negative, for $r_I \sim r_+(m_1)$, and
goes positive for $r_I \sim r_-(m_2)$.  So we are guaranteed a value
of $r_I$ for which the two expressions are identical, and we simply
take that value for which they are closest.

We can use the resulting relation $r_I(\tilde{E}_1)$ to plot the
proper length, $L$, and the time at
which the spacetime is perturbed, $t_p$, as functions of
$\tilde{E}_1$. The proper length of the geodesic in the perturbed spacetime is
\[
L (\Delta m) =
\int_{r_{\tilde{E}_1}(m_1)}^{\infty} \frac{\d r}{\sqrt{\tilde{E}^2_1 -
V_{m_1}(r)}} - \int_{r_{\tilde{E}_1}(m_1)}^{r_I} \frac{\d
r}{\sqrt{\tilde{E}^2_1 - V_{m_1}(r)}}
\]
\be
+\int_{r_{\tilde{E}_2}(m_2)}^{\infty} \frac{\d r}{\sqrt{\tilde{E}^2_2 -
V_{m_2}(r)}} + \int_{r_{\tilde{E}_2}(m_2)}^{r_I} \frac{\d
r}{\sqrt{\tilde{E}^2_2 - V_{m_2}(r)}}
\ee
and the time $t_p$ is
\[
t_p = E_1 \int_{r_{E_1}(m_1)}^\infty \frac{\d r}{V_{m_1}(r)
  \sqrt{E^2_1 - V_{m_1}(r)}} - \tilde{E}_1
\int_{r_{\tilde{E}_1}(m_1)}^\infty \frac{\d r}{V_{m_1}(r)
  \sqrt{E^2_1 - V_{m_1}(r)}}
\]
\be + \tilde{E}_1 \int_{r_{\tilde{E}_1}(m_1)}^{r_I} \frac{\d
  r}{V_{m_1}(r) \sqrt{\tilde{E}^2_1 - V_{m_1}(r)}} - r_*(r_I) + r_*(\infty).
\ee
Both expressions can, again, be evaluated in terms of elliptic
integrals as in section 3. Both $L$ and $t_p$ are functions of
$\tilde{E}_1$, utilising the relation $r_I(\tilde{E}_1)$ found
previously.

To recap, we run through values of $\tilde{E}_1$, then run through
values of $r_I$ at each step, computing $\tilde{E}_2$, and evaluating
(\ref{eqn:t1}) and (\ref{eqn:t2}).  We find a relation
$r_I(\tilde{E}_1)$ by taking those values of $r_I$ for which
(\ref{eqn:t1}) and (\ref{eqn:t2}) are approximately equal, then
evaluate $L$ and $t_p$ for each
$\{\tilde{E}_1,r_I\}$.  It is then a simple matter to plot the
function $\delta L(t_p)$ as given in (\ref{eqn:deltaL}). The result
is shown in figure 12.  This schematic plot is for some specific
values of the parameters ($x_- \sim 0.25, x_+ \sim 2.5$, which gives
$m_1 = 10$ and $Q = 1.5$, $\Delta m = 0.5$ and $E^2 \sim 0.999
V_{\rm max}(m_1)$), but it seems to be representative of the general
behaviour.

\vskip 0.5cm\begin{figure}[ht]
         \beginlabels\refpos 0 0 {}
                     \put 150 0 {\delta L}
                     \put 303 -190 {t_p}
         \endlabels
         \centerline{
         \psfig{figure=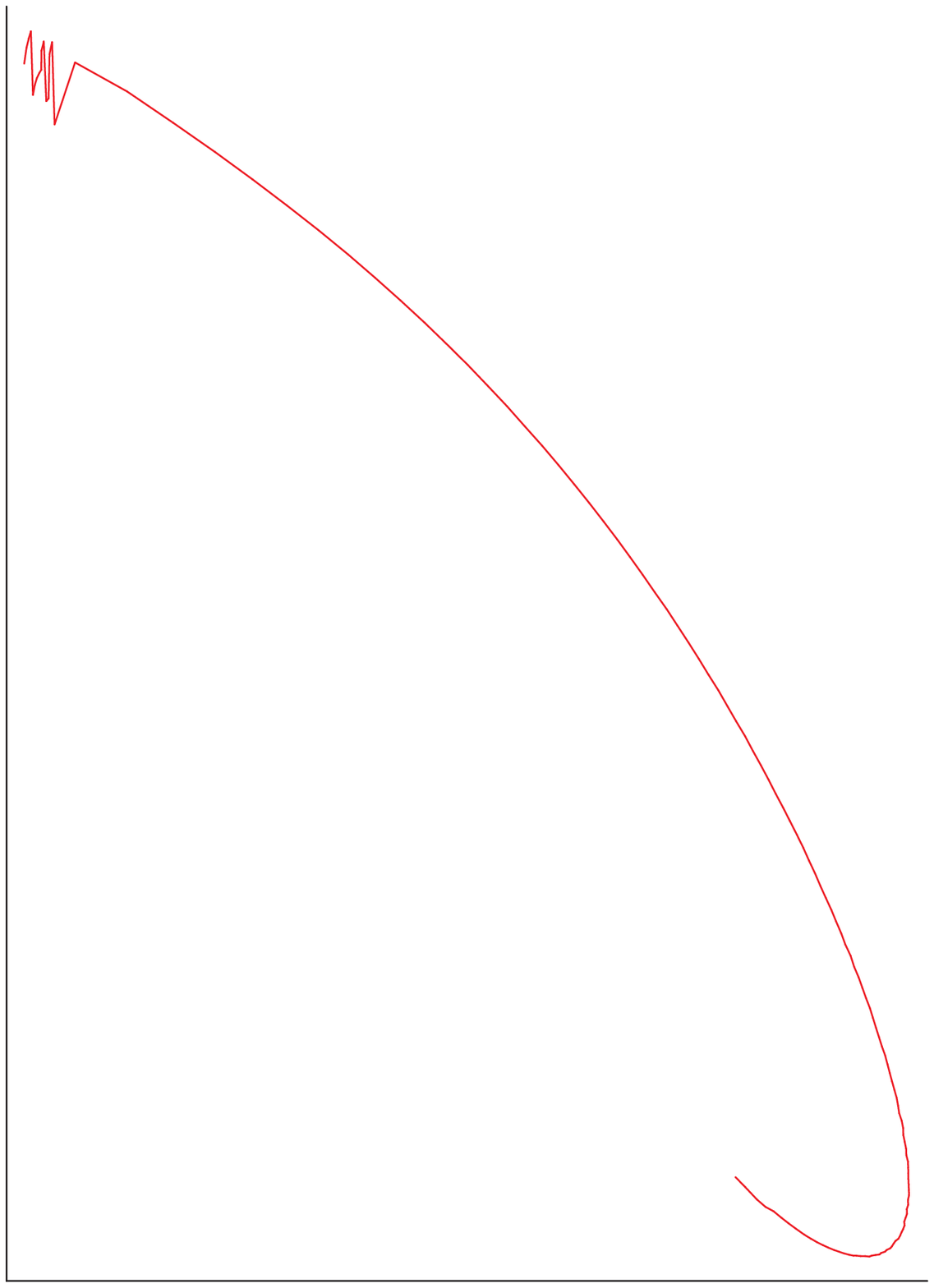,height=2.5in} }
         \vskip0.5cm
         {\footnotesize Figure 12.  Here we show the fractional change
         in the length of the geodesic as a function of $t_p$, for
         some choice of parameters as explained in the text.  The
         lack of smoothness for small $t_p$ is due to numerical effects, and is
         not expected to be physical.}
\end{figure}

This plot demonstrates that our correlators are sensitive to
a perturbation of the geometry which is localised beyond the outer
horizon. We postpone further discussion for future work, and
conclude by making a few comments.

First, we should note that figure 12 cannot be used for the region
$V_{\rm max}(m_1) < \tilde{E}^2_1 < V_{\rm max}(m_2)$ of energies, since in
this case the integrals have to be evaluated differently. Similar
methods should work to generate a plot of $\delta L(t_p)$ for that
range of energies however.

Moreover, we can only probe part of the region of the spacetime
beyond the event horizon. There are no real solutions for $r_I$
beyond a certain range of energies, and this range does not cover
the entire region behind the event horizon.

Finally, it is  intriguing that the plot of $L(t_p)$ in figure 12 is
not single-valued.  This  means that there are two geodesics, with
different energies $\tilde{E}_1$, which connect the fixed boundary
points in the perturbed spacetime. Similar to the case of the
unperturbed spacetime, this is reflected in some non-analytic
behaviour of the correlators in the perturbed spacetime. It would be
nice to get a better  handle, through analytic calculations, on these
issues.


\section*{Acknowledgments} We would like to thank Micha Berkooz,
Don Marolf, Simon Ross, Paul Saffin and Kristin Schleich for
discussions, and are grateful to Simon Ross for pointing out an error
in the initial version of this paper.  DB is supported in part
by a postdoctoral fellowship from the Pacific Institute for the
Mathematical Sciences.  MR would like to thank the Physics
Departments of Cornell and Syracuse Universities and the KITP in
Santa Barbara for hospitality during the course of this work, which
was supported in part by the Natural Sciences and Engineering Council
of Canada and the National Science Foundation under Grant No. PHY99-07949.


\section*{Appendix: The $t(E)$ Integral}

Consider the integral (\ref{relation1}) for non-rotating geodesics.
Then $V(r) = -f(r)$ and, dropping the residue coming from
integrating over the horizon, we have \be -t_0 = \frac{E}{2}
\int^{\infty}_{\td{x}_+} \d x \frac{
x^{5/2}}{(x+x_0)(x-x_-)(x-x_+)\sqrt{(x+\td{x}_0)(x-\td{x}_-)(x-\td{x}_+)}}.
\ee Defining a new variable $t$ through \be t^2 = a
\left(\frac{x-\td{x}_+}{x-\td{x}_-}\right), \ee with $a$ a constant
to be determined, gives \be -t_0 = \frac{E}{2} A \int_0^a \d t \left(
1 - \frac{\td{c}_0}{a} t^2 \right)^{-1/2} \left( 1 -
\frac{\td{x}_-}{a\td{x}_+} t^2 \right)^{5/2} \left( \left( 1 -
\frac{c_0}{a} t^2 \right) \left( 1 - \frac{c_-}{a} t^2 \right)
\left( 1 - \frac{c_+}{a} t^2 \right) \right)^{-1} \ee where we have
set \[ A = \frac{ 2\td{x}_+^{5/2} }{ a^{1/2}
(\td{x}_++\td{x}_0)^{1/2}}
\left((\td{x}_++x_0)(\td{x}_+-x_-)(\td{x}_+-x_+)\right)^{-1}, \] \be
\td{c}_0 = \frac{\td{x}_-+\td{x}_0}{\td{x}_++\td{x}_0}, \qquad c_0 =
\frac{\td{x}_-+x_0}{\td{x}_++x_0}, \qquad c_- =
\frac{\td{x}_--x_-}{\td{x}_+-x_-}, \qquad c_+ =
\frac{\td{x}_--x_+}{\td{x}_+-x_+}. \ee Now observe that \be \left(
\left( 1 - \frac{c_0}{a} t^2 \right) \left( 1 - \frac{c_-}{a} t^2
\right) \left( 1 - \frac{c_+}{a} t^2 \right) \right)^{-1} =
\hat{c}_0 \left( 1 - \frac{c_0}{a} t^2 \right)^{-1} + \hat{c}_-
\left( 1 - \frac{c_-}{a} t^2 \right)^{-1} + \hat{c}_+ \left( 1 -
\frac{c_+}{a} t^2 \right)^{-1} \ee where \be \hat{c}_0 =
\frac{c_0^2}{(c_0-c_-)(c_0-c_+)}, \qquad \hat{c}_- =
\frac{c_-^2}{(c_--c_0)(c_--c_+)}, \qquad \hat{c}_+ =
\frac{c_+^2}{(c_+-c_0)(c_+-c_-)}. \ee If we take \be a=\td{c}_0,
\qquad k^2 = \frac{\td{x}_-}{\td{x}_+} \frac{1}{a}, \qquad
\alpha_0^2 = \frac{c_0}{a}, \qquad \alpha_-^2 = \frac{c_-}{a},
\qquad \alpha_+^2 = \frac{c_+}{a}, \ee so that \be \hat{c}_0 =
\frac{\alpha_0^4}{(\alpha_0^2-\alpha_-^2)(\alpha_0^2-\alpha_+^2)},
\qquad \hat{c}_- =
\frac{\alpha_-^4}{(\alpha_-^2-\alpha_0^2)(\alpha_-^2-\alpha_+^2)},
\qquad \hat{c}_+ =
\frac{\alpha_+^4}{(\alpha_+^2-\alpha_0^2)(\alpha_+^2-\alpha_-^2)},
\ee then the integral becomes \be -t_0 = \frac{E}{2} A \int_0^a \d t
\sqrt{\frac{(1-k^2t^2)^5}{1-t^2}} \left(
\frac{\hat{c}_0}{(1-\alpha_0^2 t^2)} +
\frac{\hat{c}_-}{(1-\alpha_-^2 t^2)} +
\frac{\hat{c}_+}{(1-\alpha_+^2 t^2)}\right). \ee We now use \be
\sqrt{\frac{1-k^2t^2}{1-t^2}} \frac{1}{(1-\alpha^2 t^2)} =
\frac{(\alpha^2-k^2)}{\alpha^2}
\frac{1}{(1-\alpha^2t^2)\sqrt{(1-t^2)(1-k^2t^2)}} +
\frac{k^2}{\alpha^2} \frac{1}{\sqrt{(1-t^2)(1-k^2t^2)}} \ee and,
noting that \be
\frac{\hat{c}_0}{\alpha_0^2}+\frac{\hat{c}_-}{\alpha^2_-}+\frac{\hat{c}_+}{\alpha^2_+}=
0, \ee we have \be -t_0 = \frac{E}{2} A \int_0^a \d t \frac{(1-k^2
 t^2)^2}{\sqrt{(1-t^2)(1-k^2t^2)}} \left( I_0 + I_- +
 I_+ \right)
\ee where
\begin{eqnarray}
I_0 &=& \frac{ \alpha_0^2 (\alpha_0^2-k^2) }{
(\alpha_0^2-\alpha_-^2)(\alpha_0^2-\alpha_+^2) }
~\frac{1}{
(1-\alpha_0^2t^2) }, \\
I_- &=& \frac{ \alpha_-^2 (\alpha_-^2-k^2) }{
(\alpha_-^2-\alpha_0^2)(\alpha_-^2-\alpha_+^2) }
~\frac{1}{
(1-\alpha_-^2t^2) }, \\
I_+ &=& \frac{ \alpha_+^2 (\alpha_+^2-k^2) }{
(\alpha_+^2-\alpha_0^2)(\alpha_+^2-\alpha_-^2) }
~\frac{1}{ (1-\alpha_+^2t^2) }.
\end{eqnarray}
Each term can now be evaluated~\cite{elliptic}, to give \[ -t_0 =
\frac{E}{2} A \left[ \frac{k^6}{\alpha_0^2\alpha_-^2\alpha_+^2} {\rm
F}(\p,k) +
\frac{(\alpha_0^2-k^2)^3}{\alpha_0^2(\alpha_0^2-\alpha_-^2)(\alpha_0^2-\alpha_+^2)}
\Pi(\p,\alpha_0^2,k) \right. \]
\be+
\left.
\frac{(\alpha_-^2-k^2)^3}{\alpha_-^2(\alpha_-^2-\alpha_0^2)(\alpha_-^2-\alpha_+^2)}
\Pi(\p,\alpha_-^2,k) +
\frac{(\alpha_+^2-k^2)^3}{\alpha_+^2(\alpha_+^2-\alpha_0^2)(\alpha_+^2-\alpha_-^2)}
\Pi(\p,\alpha_+^2,k) \right]
\ee
where $\p=\sin^{-1}a$.  Substituting for the
various constants gives the result (\ref{eqn:time})
in the text.


\end{document}